# The disclosure of information about the range of asset value in market

Jianhao Su
sujianhao_1@sdnu.edu.cn
Shandong Normal university

Yanliang Zhang
zhangyl@sdufe.edu.cn
Shandong University of Finance and Economics

## Abstract

The information released to investors in financial markets has various forms. We refer to range information as information about the upper and lower bound which the payoff of a risky asset may reach in the future. This study develops rational expectation models to explore the market impacts of disclosure of range information. Our model shows that its disclosure can decrease the sensitivity of market price to private signal and increase market liquidity. The market impact of its disclosure depends on the position and precision of the disclosed range. When the linear combination of private signal and noise trading volume is distant from the disclosed range, the reaction of price to a variation in private signal will almost vanish, whereas a movement of the disclosed range can affect the price efficiently. Under certain conditions, such as a higher proportion of informed traders or a larger size of noise trading in market, the disclosure of range information is more likely to reduce asset price and raise capital cost.

## 1.Introduction

In financial market, the completeness of information held by different investors varies, which is known as information asymmetry among investors. The information disclosure to market investors has a great impact on the operation of market. A large number of literature specifically focuses on the impact of information distribution on the formation of market price, the sensitivity of price to released information and market quality. Market quality mainly includes market liquidity, price volatility, market efficiency, etc.

A large part of the theoretical research on this topic is developed under the framework of rational expectation. This series of studies assumes that traders with less information can use market price to infer the private information held by traders who have more information, and apply the outcomes of their inference to make investment decisions.

Grossman and Stigliz (1980) [1] is an early study based on the rational expectation framework. In its setup, there are informed traders and uninformed traders in the market. Informed traders can obtain a private signal about asset value, while uninformed traders can only observe asset price and then infer the private signal of informed traders by the price. Under this setup, Grossman and Stigliz (1980) [1] analyzed the trading behavior of the two types of traders and market equilibrium.

In recent years, research literature on this topic has focused more on investors' acquisition and processing of information, as well as the impact of their information acquisition and processing behavior on market prices and quality. Also based on the framework of rational expectation model, this series of literature argues that investors' information learning and processing can be more complex in some situations, which changes their trading behavior. These situations mainly include the following three.

(1) Investors face ambiguity about some parameters of asset markets.

(2) Asset value is composed of multiple fundamentals.

(3) There is a deviation in investors' interpretation of information in market.

The first strand of literature is developed under the situation that some investors' belief about some key parameters is ambiguous, which makes it more complicated for them to confirm the optimal trading volume. For example, Epstein and Schneider (2008) [2] discussed the market where some investors have ambiguity in their perception of the accuracy of signal. Mele and Sangiorgi (2015) [3], Hahn and Kwon (2015) [4], Condie and Ganguli (2017) [5], and Huo et al.(2024) [6] focused on the market where there is ambiguity in some investors' belief about the unconditional mean of asset value. Easley et al. (2014) [7], Huang et al. (2017) [8], Illeditsch et al. (2021) [9] considered the market where some investors are ambiguous about the correlation coefficient between different asset values. Anthropelos and Schneider (2024) [10], Ghazi et al. (2024) [11] introduce ambiguity into the distribution of asset payoff in investors' belief. These studies show that the ambiguity of investors can lead to significant changes in their trading behavior, market price and quality.

The second strand of literature analyzes the market where asset value is determined by multiple fundamentals and different groups of informed traders can obtain private signals about different fundamentals, such as Kondor (2012) [12], Goldtein and Yang (2015) [13], Liu et al. (2019) [14], Yang and Zhu (2020) [15]. When various traders participate in transactions, the information about multiple fundamentals will be injected into market price simultaneously and these fundamentals will be reflected in asset price in a complicated form, which can make it more difficult for uninformed traders to infer the information about asset fundamentals from the price and reducing the accuracy of their inference results. The above literature shows that investors’ trading behavior, market price and quality will be distinct under the setup of multiple fundamentals.

The third strand of literature is developed under the situation that there is a deviation in some investors' interpretation of the signals about asset in market, which can make their trading decision deviate from the optimal. They analyzes how equilibrium price and market quality would be affected by the deviation, such as Dudams et al. (2017) [16], Banerjee et al. (2018) [17], Liu et al. (2018) [18], and Hu and Wang (2024) [19].

Existing studies usually depict the information acquired by investors in two similar ways. Suppose risky asset has an uncertain future value $v$. Based on the assumption that $v$ consists of several parts, such as $v = v_1 + v_2$, the first way is to take the realized value of $v_1$ or $v_2$ as the information acquired by investors.

Use $\eta$ to represent a stochastic error item. The second way is to take the realized value of $v + \eta$ as the information acquired by investors.

The two ways of depicting the information are essentially the same and we refer to them as common-form signals. For reflecting the information asymmetry in market, related literature usually assumes that informed traders can receive this form of signals while uninformed traders cannot, or informed traders can receive more signals than uninformed traders.

However, there are more forms of information that investors can obtain in market. Some information can reflect the upper or lower bound that the future value of assets may reach. We refer to this type of information as the range information about asset value.

For example, many countries have a price limit regime in their securities markets, under which investors tend to believe that the future value of a security must have an upper and lower bound.

Additionally, security analysts are often keen on making predictions about the upper or lower bound of asset value in future and their predictions can possibly convince investors.

In some cases, a public company or some financial institutions need prevent the stock price from falling below a certain level and they can take some powerful measures to stop the decline of its stock price under the condition that the price falls to the level. Similarly, some block-holders may sell their shares in large quantities to realize profits when the stock price rises to a certain level they previously set, which can prevent the price from rising further. Once the insider information above are acquired by investors, they will also form the belief about the upper and lower bound of the asset value in the future.

In summary, public companies, financial institutions, security analysts, block-holders or governments may intervene in asset market price, which can cause its price has a lower or upper bound and then limit its future value in a range. In reality, the information about the series of interventions should be regarded as range information and some of this type of information can be disclosed to the public.

Thus, range information is likely to exist in investors’ information set and affects their belief. The main issue we want to explore is the impact of disclosure of range information on investors' trading behavior, the formation of market price, and market quality.

Range information has not been studied and introduced into models of market micro-structure by existing theoretical research. By introducing range information into rational expectation models, this study reveals that disclosure of range information can distinctly changes investors' behavior, market equilibrium and market quality compared with existing theoretical literature. To be concrete, we reveal whether market price and market quality (such as liquidity and asset premium) will increase or decrease after range information is disclosed to investors, and how market price reacts to a variation in private signal and the disclosed range when range information is disclosed to investors. For example, we find that its disclosure can cause that market liquidity and the reaction of market price to private signal will vary continuously with private signal and noise trading volume. In the models of related literature, market liquidity and the reaction of price is usually constant or piecewise constant with private signal and noise trading. So we reveal that its disclosure would lead to a new pattern of market price reacting to private signal and noise trading, and a new pattern of market liquidity varying with them, which is rarely recorded by existing literature.

In the model of Grossman and Stigliz (1980) [1], all investors explicitly know the probability distribution of asset value. The first strand of related literature introduces ambiguity into some investors' perception of the parameters of the distribution. Thus, they don not explicitly know the distribution of asset value, only that it belongs to a parametric distribution family, which implies the introduction of ambiguity weakens these investors' information set. The second strand of related literature introduces multiple fundamentals into the composition of asset value. Under this setup, the asset value is composed of several mutually independent fundamentals but none of their realized value can be acquired by uninformed traders. Thus, the introduction of multiple fundamentals also weakens these investors' information set. The third strand of related literature introduces deviation into some investors' interpretation of the signal about asset value. This results in a bias in the information acquired by them relative to the actual signal and then also weakens their information set. However, the disclosure of range information would increase the information acquired by investors and they need to make decision based on more types of information. Thus, in contrast to these existing studies, investors' information set is strengthened in our model.

The remainder of this paper is organized as follows. Section 2 presents our main model by which we analyze the trading behavior of various investors and the market equilibrium when range information is disclosed to investors. In section 3, we use a benchmark model to present the trading behavior of various traders and market equilibrium in the situation where range information is not disclosed to investors. Section 4 analyzes market price, the reaction of price to variation in private signal and the disclosed range, market liquidity, asset premium in the equilibrium of our main model and compares the equilibriums of the two models to reveal the impact of range information disclosure. Finally, section 5 concludes.

## 2.The Main Model

## 2.1 Setup

Following Mondria et al. (2022) [20] and Huang et al. (2020) [21], we assume that asset market is perfectly competitive and lasts for three periods: t=0, 1 and 2.

There are two assets traded in the market. The first asset is a risk-free bond which is unlimited supply. Following Easley et al. (2014) [7] and Mondria et al. (2022) [20], we also suppose for simplicity that its price is always 1 at three periods. The second asset is the stock of a public company which has a total supply of $Z$ units. It should be regarded as a risky asset.

Use $v$ to denote the payoff on one unit of risky asset. At t=0, $v$ is uncertain for all investors and can be regarded as the future value of the risky asset, which means nobody knows the realized value of $v$. As time goes by, the realized value of $v$ is revealed and observable to all investors at t=1. At t=2, the risky asset pays off its holders and a holder of each unit can obtain an amount of $v$.

Investors are all short-term traders and exist in the market for only two periods. More concretely, the first generation of investors trade at t=0 and have to sell their assets in market to get cash for their consumption at t=1. Similarly, the second generation of investors trade at t=1 and get the payoffs of their assets at t=2.

Following Grossman and Stiglitz (1980) [1] and Huang et al. (2020) [21], we assume that

$$v = u + \varepsilon,$$

where $u \sim N(\mu_0, \sigma_u^2)$ and $\varepsilon \sim N(0, \sigma_\varepsilon^2)$. $u$ and $\varepsilon$ are mutually independent. $u$ can be regarded as the fundamental of the company and $\varepsilon$ can be referred to as the disturbance of some unobservable random factors on its future value.

At t=0, investors can finance their purchase of risky assets by selling risk-free bonds short, and there are no restrictions on the amount of short selling.

Following Goldstein and Yang (2017) [22], there are three types of investors in the market: informed traders, uninformed traders and noise traders. Suppose the total mass of informed and uninformed traders is 1. The fraction of informed traders and uniformed traders are respectively $x_I$ and $x_U$ with $x_I + x_U = 1$.

Noise traders totally demand $y$ units of the risky asset at t=0, that is, $y$ is noise trading volume. $y \sim N(0, \sigma_y^2)$ and $y$ is independent of $u$ and $\varepsilon$. A larger $\sigma_y^2$ implies a greater size of noise trading in the market. In the following, we use $\tilde{y}$ to denote the realized value of $y$.

At t=0, informed traders can observe the realized value of $u$, while uninformed traders cannot, which implies $u$ can be regarded as informed traders' private signal about $v$. A larger value of its variance $\sigma_u^2$ implies a greater informativeness of the private signal. $\varepsilon$ is unobservable to all investors and can be treated as the error of informed traders' private signal $u$. Given that $Var[v|u] = \sigma_\varepsilon^2$, a smaller value of its variance $\sigma_\varepsilon^2$ implies a smaller deviation of $u$ from the asset future value $v$ and then a more precise signal received by informed traders.

Related literature typically assumes that uninformed traders are aware that informed traders' information are more sufficient than theirs, such as Goldstein and Yang (2015) [13], Easley and O'Hara (2009) [23]. We adhere to this assumption by supposing that uninformed traders perceive that informed traders can receive a private signal $u$ while they cannot.

Next, we introduce range information about risky asset value at t=1 into our setup as follows.

There is a price limit regime in the market and the stock cannot be traded at a price lower than $\underline{v}_r$ or high than $\overline{v}_r$ at t=1 under this regime. In addition, the management of the public company needs to keep its stock price above a certain level for some reasons, such as ensuring the stability of their positions or preventing the company from being delisted. Then, in order to boost market confidence, the company at t=0 promises to provide its shareholders with a stock repurchase offer at t=1 with an offer price of $\underline{v}_c$, where $\underline{v}_c > \underline{v}_r$ and $\underline{v}_c < \overline{v}_r$ .

Suppose that the range information above is disclosed to all investors at t=0. For the first generation of investors, if the payoff $v$ revealed at t=1 is between $\underline{v}_c$ and $\overline{v}_r$, then they can sell their stocks in market at a price of $v$ given the full competitiveness of the market. If $v$ revealed at t=1 is higher than $\overline{v}_r$, then they still have to sell their stocks at a price of $\overline{v}_r$ for their consumption. If $v$ revealed at t=1 is lower than $\underline{v}_c$, then they would accept the repurchase offer and sell their stocks to the company at a price of $\underline{v}_c$ per share.

As a result, the disclosure of the range information at t=0 suggests that the upper and lower bound of the risky asset value at t=1 are respectively $\overline{v}_r$ and $\underline{v}_c$, that is, $v \in [\underline{v}, \overline{v}]$, where $\underline{v} = \underline{v}_c$ and $\overline{v} = \overline{v}_r$.

Before trading, every investor is endowed with $D_0$ units of the risk-free bond.

At t=0, the price of the risky asset is denoted by $p$. Use $\theta_I$ and $\theta_U$ to denote informed traders' and uninformed traders' demand for the risky asset respectively. Then their wealth at t=1 are respectively

$$W_I = D_0 + \theta_I(v - p)$$

and

$$W_U = D_0 + \theta_U(v - p).$$

Suppose that all the investors have a CARA utility function with a common absolute risk aversion coefficient $\gamma$, that is,

$$U(W) = -e^{-\gamma \cdot W},$$

where $W$ is the wealth held by the investor at t=1. At t=0, every investor trade in order to maximize the expectation of his utility conditional on his information set.

Under our assumptions, the market clearing condition should be

$$x_I\theta_I + x_U\theta_U + \tilde{y} = Z. \qquad (1)$$

## 2.2 The decision of informed traders

If the price of the risky asset $p \geq \overline{v}$, selling the asset short will be profitable with probability 1 and then informed traders are bound to do that infinitely. If $p \leq \underline{v}$, informed traders are bound to buy the asset infinitely. So, only when $\underline{v} < p < \overline{v}$ can the market reach equilibrium. The premise of our following discussion is $p \in (\underline{v}, \overline{v})$.

Now consider a representative informed trader. The realized value of $u$ is denoted by $\tilde{u}$, and his information set is $\{u = \tilde{u},\ v \in [\underline{v}, \overline{v}]\}$. The conditional expectation of

his utility is

$$U_I(\theta_I) \triangleq E\left[-e^{-\gamma\left(D_0+\theta_I(v-p)\right)}\middle| u=\tilde{u}, v\in\left[\underline{v}, \overline{v}\right]\right].$$

His decision problem is

$$\max_{\theta_I} \; U_I(\theta_I).$$

Use $\Psi(\cdot)$ and $\psi(\cdot)$ to denote the distribution and density function of the standard normal distribution respectively. We can derive

$$U_I(\theta_I) = \frac{E\left[-e^{-\gamma\left(D_0+\theta_I(v-p)\right)}\mathbf{1}_{\{v\in[\underline{v},\overline{v}]\}}| u=\tilde{u}\right]}{E\left[\mathbf{1}_{\{v\in[\underline{v},\overline{v}]\}}| u=\tilde{u}\right]}$$

$$= -e^{-\gamma D_0+\gamma(p-\tilde{u})\theta_I+\frac{\gamma^2\sigma_\varepsilon^2\theta_I^2}{2}} \cdot \frac{\Psi\left(\frac{\overline{v}+\gamma\sigma_\varepsilon^2\theta_I-\tilde{u}}{\sigma_\varepsilon}\right)-\Psi\left(\frac{\underline{v}+\gamma\sigma_\varepsilon^2\theta_I-\tilde{u}}{\sigma_\varepsilon}\right)}{\Psi\left(\frac{\overline{v}-\tilde{u}}{\sigma_\varepsilon}\right)-\Psi\left(\frac{\underline{v}-\tilde{u}}{\sigma_\varepsilon}\right)}. \quad (2)$$

We give the details about the calculation of equation (2) in appendix A.1. The derivative of the utility with respect to $\theta_I$ is

$$U_I{}'(\theta_I) = -\gamma \cdot U_I(\theta_I) \cdot \left[\boldsymbol{J}_{[\underline{v},\ \overline{v}]}(\tilde{u}-\gamma\sigma_\varepsilon^2\theta_I)-p\right], \quad (3)$$

where $\boldsymbol{J}_{[\underline{v},\ \overline{v}]}(t) \triangleq \sigma_\varepsilon\left[\frac{\psi\left(\frac{\underline{v}-t}{\sigma_\varepsilon}\right)-\psi\left(\frac{\overline{v}-t}{\sigma_\varepsilon}\right)}{\Psi\left(\frac{\overline{v}-t}{\sigma_\varepsilon}\right)-\Psi\left(\frac{\underline{v}-t}{\sigma_\varepsilon}\right)}\right]+t.$

We give a closed form solution of an informed trader's optimal decision by the following proposition. The proof is given in appendix A.3.

**Proposition 1.** For a given price $p\in(\underline{v},\ \overline{v})$, an informed trader's optimal demand for the risky asset is $\widetilde{\theta_I} = \frac{1}{\gamma\sigma_\varepsilon^2}\cdot\tilde{u}+k$, where the parameter $k$ is the solution of

$$p = \boldsymbol{J}_{[\underline{v},\ \overline{v}]}(-\gamma\sigma_\varepsilon^2 k). \quad (4)$$

The equation above has a unique solution. $\widetilde{\theta_I}$ decreases monotonically with the price $p$.

According to the item (3) of appendix A.2, we have $\boldsymbol{J}_{[\underline{v},\ \overline{v}]}{}'(t)>0$, which implies $\boldsymbol{J}_{[\underline{v},\ \overline{v}]}(t)$ is continuous and increases monotonically with $t$. So it must have an inverse function $\boldsymbol{J}^{-1}_{[\underline{v},\ \overline{v}]}(\cdot)$. Proposition 1 actually shows that the optimal demand is

$$\widetilde{\theta_I} = \frac{1}{\gamma\sigma_\varepsilon^2}\cdot\tilde{u}-\frac{\boldsymbol{J}^{-1}_{[\underline{v},\ \overline{v}]}(p)}{\gamma\sigma_\varepsilon^2},$$

which implies $\widetilde{\theta_I}$ includes two parts. The first part $\frac{1}{\gamma\sigma_\varepsilon^2}\cdot\tilde{u}$ depends on the private signal $\tilde{u}$ and is independent of the range information (i.e., the lower bound $\underline{v}$ and the upper bound $\overline{v}$) and the price $p$. On the contrary, the second part $-\frac{J^{-1}_{[\underline{v},\ \overline{v}]}(p)}{\gamma\sigma_\varepsilon^2}$ depends on the range information and the price $p$, but cannot be affected by the private signal $\tilde{u}$.

### 2.3 The decision of uninformed traders

Remind uniformed traders cannot receive the private signal $\tilde{u}$, but can observe the price $p$ and range information $[\underline{v},\overline{v}]$.

Informed traders inject their private signal into the price when they trade in the market. Thus, related literature under the framework of rational expectation usually assumes that uninformed traders try to infer the private signal from the price. In our model, the price is also impacted by the disclosed range $[\underline{v},\overline{v}]$ because informed traders' trading volume depends on $\underline{v}$ and $\overline{v}$ by Proposition 1.

Before analyzing uninformed traders' inference, related literature always conjectures the equilibrium price $p$, setting $p$ in a linear form of private signal and noise trading volume, such as Easley et al.(2014) [7], Goldstein and Yang(2017) [22]. However, under our setup, $p$ must be between $\underline{v}$ and $\overline{v}$, which implies that it cannot be set in a linear form. According to equation (4), we conjecture the equilibrium price as a quasi-linear form of private signal and noise trading volume

$$p = \boldsymbol{J}_{[\underline{v},\ \overline{v}]}(\tau\tilde{u} + \alpha\tilde{y} + \beta), \tag{5}$$

where the coefficients $\tau$,$\alpha$, $\beta$ will be determined in equilibrium. According to the item (1) of appendix A.2, if $p$ is given by equation(5), it follows that $p \in (\underline{v},\ \overline{v})$.

By Proposition 1, under the price given by equation(5), we have $k = -\frac{\tau}{\gamma\sigma_\varepsilon^2}\tilde{u} - \frac{\alpha}{\gamma\sigma_\varepsilon^2}\tilde{y} - \frac{\beta}{\gamma\sigma_\varepsilon^2}$ and an informed trader' demand is

$$\widetilde{\theta_I} = \frac{1-\tau}{\gamma\sigma_\varepsilon^2}\tilde{u} - \frac{\alpha}{\gamma\sigma_\varepsilon^2}\tilde{y} - \frac{\beta}{\gamma\sigma_\varepsilon^2}. \tag{6}$$

For a representative uninformed trader, he knows the relationship between the price and private signal is depicted as equation(5) and then rationally infers that

$$\tilde{u} + \frac{\alpha}{\tau}\tilde{y} = \frac{\boldsymbol{J}^{-1}_{[\underline{v},\overline{v}]}(p) - \beta}{\tau}.$$

By the above equation, it can be seen that his inference about the private signal $\tilde{u}$ is polluted by the noise trading $\tilde{y}$ and varies with $\underline{v}$ and $\overline{v}$.

In this case, the conditional expectation of his utility at t=1 should be

$$U_U(\theta_U;\underline{v},\overline{v}) \triangleq E\left[-e^{-\gamma(D_0+\theta_U(v-p))}\middle| u + \frac{\alpha}{\tau}y = \frac{\boldsymbol{J}^{-1}_{[\underline{v},\overline{v}]}(p) - \beta}{\tau}, v \in [\underline{v},\overline{v}]\right].$$

By some calculations, we derive

$$U_U\left(\theta_U;\underline{v},\overline{v}\right)=-e^{\gamma(p-\mu_\eta)\theta_U+\frac{\gamma^2\sigma_\eta^2\theta_U^2}{2}}$$

$$\cdot\left[\frac{\Psi\left(\frac{\overline{v}+\gamma\sigma_\eta^2\theta_U-\mu_\eta}{\sigma_\eta}\right)-\Psi\left(\frac{\underline{v}+\gamma\sigma_\eta^2\theta_U-\mu_\eta}{\sigma_\eta}\right)}{\Psi\left(\frac{\overline{v}-\mu_\eta}{\sigma_\eta}\right)-\Psi\left(\frac{\underline{v}-\mu_\eta}{\sigma_\eta}\right)}\right], \tag{7}$$

where $\mu_\eta=\omega_1\mu_0+\omega_2\left(\frac{\boldsymbol{J}_{[\underline{v},\overline{v}]}^{-1}(p)-\beta}{\tau}\right)$, $\sigma_\eta^2=\sigma_\varepsilon^2+\omega_1\sigma_u^2$ with $\omega_1=\frac{\alpha^2\sigma_y^2}{\tau^2\sigma_u^2+\alpha^2\sigma_y^2}$, $\omega_2=1-\omega_1=\frac{\tau^2\sigma_u^2}{\tau^2\sigma_u^2+\alpha^2\sigma_y^2}$. The details about the calculation of equation (7) can be seen in appendix A.4.

As for the optimal decision of the uninformed trader, we have the following proposition. The proof is given in appendix A.5.

**Proposition 2**. $U_U\left(\theta_U;\underline{v},\overline{v}\right)$ reaches its maximum at

$$\overline{\theta}_U=\frac{1}{\gamma\sigma_\eta^2}\left[\omega_1\mu_0-\frac{\omega_2\beta}{\tau}+\left(\frac{\omega_2}{\tau}-1\right)\cdot\boldsymbol{J}_{[\underline{v},\overline{v}]}^{-1}(p)\right]. \tag{8}$$

Additionally, $U_U\left(\theta_U;\underline{v},\overline{v}\right)$ increases strictly in $(-\infty,\overline{\theta}_U)$ and decreases strictly in $(\overline{\theta}_U,+\infty)$.

As a result, an uninformed trader's optimal demand is $\overline{\theta}_U$, which also depends on the lower bound $\underline{v}$ and the upper bound $\overline{v}$.

## 2.4 The market equilibrium

In the market equilibrium, the demand of an uninformed trader and an informed trader are given by equation (8) and (6), respectively.

Insert their trading volume into the market clearing condition (i.e., equation (1)), we derive

$$x_I\left(\frac{1-\tau}{\gamma\sigma_\varepsilon^2}\tilde{u}-\frac{\alpha}{\gamma\sigma_\varepsilon^2}\tilde{y}-\frac{\beta}{\gamma\sigma_\varepsilon^2}\right)+\frac{x_U}{\gamma\sigma_\eta^2}\left[\omega_1\mu_0-\frac{\omega_2\beta}{\tau}+\left(\frac{\omega_2}{\tau}-1\right)\cdot\boldsymbol{J}_{[\underline{v},\overline{v}]}^{-1}(p)\right]+\tilde{y}=Z.$$

By equation (5), we have $\boldsymbol{J}_{[\underline{v},\overline{v}]}^{-1}(p)=\tau\tilde{u}+\alpha\tilde{y}+\beta$ and inserting it into the equation above gives rise to

$$\left[x_I\left(\frac{1-\tau}{\gamma\sigma_\varepsilon^2}\right)+\frac{\tau x_U}{\gamma\sigma_\eta^2}\left(\frac{\omega_2}{\tau}-1\right)\right]\tilde{u}+\left[-\frac{\alpha x_I}{\gamma\sigma_\varepsilon^2}+\frac{\alpha x_U}{\gamma\sigma_\eta^2}\left(\frac{\omega_2}{\tau}-1\right)+1\right]\tilde{y}$$

$$+\left[-\frac{\beta x_I}{\gamma\sigma_\varepsilon^2}+\frac{x_U}{\gamma\sigma_\eta^2}(\omega_1\mu_0-\beta)\right]=Z.$$

Comparing the coefficients of $\tilde{u}$, $\tilde{y}$ and the intercept term on the left-hand side with those on the right-hand side respectively, we obtain a system of equations:

$$\begin{cases} x_I\left(\dfrac{1-\tau}{\gamma\sigma_\varepsilon^2}\right)+\dfrac{\tau x_U}{\gamma\sigma_\eta^2}\left(\dfrac{\omega_2}{\tau}-1\right)=0, \\ -\dfrac{\alpha x_I}{\gamma\sigma_\varepsilon^2}+\dfrac{\alpha x_U}{\gamma\sigma_\eta^2}\left(\dfrac{\omega_2}{\tau}-1\right)+1=0, \\ -\dfrac{\beta x_I}{\gamma\sigma_\varepsilon^2}+\dfrac{x_U}{\gamma\sigma_\eta^2}(\omega_1\mu_0-\beta)=Z. \end{cases}$$

Its solution is

$$\begin{cases} \tau = 1-\dfrac{x_U}{1+\dfrac{x_I^2\sigma_u^2}{\gamma^2\sigma_\varepsilon^4\sigma_y^2}+\dfrac{x_I\sigma_u^2}{\sigma_\varepsilon^2}}, \\ \alpha=\dfrac{\gamma\sigma_\varepsilon^2}{x_I}\cdot\tau, \\ \beta=\left(\dfrac{x_U\mu_0\omega_1}{\gamma\sigma_\eta^2}-Z\right)\Bigg/\left(\dfrac{x_U}{\gamma\sigma_\eta^2}+\dfrac{x_I}{\gamma\sigma_\varepsilon^2}\right)=\dfrac{x_U\gamma^2\sigma_\varepsilon^4\sigma_y^2(\mu_0-Z\gamma\sigma_u^2)}{{x_I}^2\sigma_u^2+\gamma^2\sigma_\varepsilon^4\sigma_y^2+x_I\gamma^2\sigma_\varepsilon^2\sigma_u^2\sigma_y^2}-Z\gamma\sigma_\varepsilon^2. \end{cases} \tag{9}$$

Therefore, the equilibrium price is

$$p_1=\boldsymbol{J}_{[\underline{v},\ \overline{v}]}(\tau\tilde{u}+\alpha\tilde{y}+\beta), \tag{10}$$

where the coefficients $\tau,\ \alpha,\ \beta$ is given by equation (9). More concretely,

$$p_1=\sigma_\varepsilon\cdot\left[\frac{\psi\left(\dfrac{\underline{v}-(\tau\tilde{u}+\alpha\tilde{y}+\beta)}{\sigma_\varepsilon}\right)-\psi\left(\dfrac{\overline{v}-(\tau\tilde{u}+\alpha\tilde{y}+\beta)}{\sigma_\varepsilon}\right)}{\Psi\left(\dfrac{\overline{v}-(\tau\tilde{u}+\alpha\tilde{y}+\beta)}{\sigma_\varepsilon}\right)-\Psi\left(\dfrac{\underline{v}-(\tau\tilde{u}+\alpha\tilde{y}+\beta)}{\sigma_\varepsilon}\right)}\right]+(\tau\tilde{u}+\alpha\tilde{y}+\beta). \tag{11}$$

Therefore, $(\tilde{\theta}_I,\bar{\theta}_U,p_1)$ constitutes a market equilibrium in the case that the range information $v\in[\underline{v},\overline{v}]$ is disclosed to investors.

Equation (11) shows that the equilibrium price $p_1$ can be impacted by the range information disclosed to informed traders, that is, $p_1$ depends on the upper bound $\overline{v}$ and the lower bound $\underline{v}$.

By the item (1) in appendix A.2, it follows that $\underline{v}<p_1<\overline{v}$, that is, the equilibrium price must be between the lower and upper bound disclosed to investors. By the item (3) in appendix A.2,

$$\frac{\partial p_1}{\partial\tilde{u}}=\tau\cdot\boldsymbol{J}_{[\underline{v},\ \overline{v}]}'(\tau\tilde{u}+\alpha\tilde{y}+\beta)>0$$

and

$$\frac{\partial p_1}{\partial\tilde{y}}=\alpha\cdot\boldsymbol{J}_{[\underline{v},\ \overline{v}]}'(\tau\tilde{u}+\alpha\tilde{y}+\beta)>0.$$

So, the equilibrium price increases with the private signal $\tilde{u}$ and noise trading volume $\tilde{y}$.

By the item (4) in appendix A.2, we have

$$\lim_{\tilde{u}\to+\infty} p_1 = \overline{v},\ \lim_{\tilde{u}\to-\infty} p_1 = \underline{v}$$

and

$$\lim_{\tilde{y}\to+\infty} p_1 = \overline{v},\ \lim_{\tilde{y}\to-\infty} p_1 = \underline{v},$$

which means the price will approach the upper bound when the private signal becomes very good (i.e., $\tilde{u}$ is very high) or the demand of noise traders becomes very large, and the price will approach the lower bound when the private signal becomes very bad (i.e., $\tilde{u}$ is very low) or the selling volume of noise traders becomes very large.

## 3. The baseline model

We give the market equilibrium under the case that there is not any range information disclosed to investors as a baseline, which has been discussed sufficiently by existing literature.

Suppose that all investors cannot obtain any range information, but informed traders can still receive the private signal $\tilde{u}$. Then the decision problem faced by a representative informed trader at t=0 is

$$\max_{\theta_I} E\left[-e^{-\gamma(D_0+\theta_I(v-p))}|u=\tilde{u}\right].$$

Its solution is

$$\theta_I = \frac{\tilde{u}-p}{\gamma\sigma_\varepsilon^2}. \tag{12}$$

Following Goldstein and Yang(2017) [22], we conjecture the equilibrium price as a linear form

$$p = \tau_0\tilde{u} + \alpha_0\tilde{y} + \beta_0.$$

Inserting it into equation (12) can derive

$$\theta_I = \frac{(1-\tau_0)\tilde{u} - \alpha_0\tilde{y} - \beta_0}{\gamma\sigma_\varepsilon^2}. \tag{13}$$

Uninformed traders can infer the private signal from the price $p$ and get

$$\tilde{u} + \frac{\alpha_0}{\tau_0}\tilde{y} = \frac{p-\beta_0}{\tau_0}.$$

Thus, for a representative uninformed trader, the decision problem is

$$\max_{\theta_U} E\left[-e^{-\gamma(D_0+\theta_U(v-p))}\,\middle|\,u + \frac{\alpha_0}{\tau_0}y = \frac{p-\beta_0}{\tau_0}\right]$$

$$= \max_{\theta_U}\ -e^{-\gamma D_0-\gamma(\mu_\eta-p)\theta_U+\frac{\gamma^2\sigma_\eta^2}{2}\cdot\theta_U^2}$$

where $\mu_\eta = \omega_1\mu_0 + \omega_2\left(\frac{p-\beta_0}{\tau}\right), \sigma_\eta^2 = \sigma_\varepsilon^2 + \omega_1\sigma_u^2$ with $\omega_1 = \frac{{\alpha_0}^2\sigma_y^2}{{\tau_0}^2\sigma_u^2+{\alpha_0}^2\sigma_y^2}$, $\omega_2 = 1-\omega_1 = \frac{\tau^2\sigma_u^2}{{\tau_0}^2\sigma_u^2+{\alpha_0}^2\sigma_y^2}$ . The solution is

$$\theta_U = \frac{\mu_\eta - p}{\gamma\sigma_\eta^2}$$

$$= \frac{1}{\gamma\sigma_\eta^2}\left[\omega_1\mu_0 + (\omega_2 - \tau_0)\tilde{u} + \left(\frac{\omega_2}{\tau_0} - 1\right)\alpha_0\tilde{y} - \beta_0\right]. \quad (14)$$

Substituting equation(13) and (14) into the market clearing condition and comparing the coefficients of $\tilde{u}$, $\tilde{y}$ and the intercept term on the left-hand side with those on the right-hand side respectively, we can derive

$$\tau_0 = \tau, \quad \alpha_0 = \alpha, \quad \beta_0 = \beta,$$

where $\tau$, $\alpha$, $\beta$ are specified in equation (9).

In summary, without any release of range information, the equilibrium price is

$$p_0 = \tau\tilde{u} + \alpha\tilde{y} + \beta, \quad (15)$$

where the coefficients $\tau$, $\alpha$, $\beta$ is given by equation (9).

## 4.The comparison between the two equilibriums

Our main model is developed under the setup that range information is disclosed to investors, whereas the baseline model is developed under the assumption that there is no disclosure of range information in market. Therefore, the comparison between the equilibriums of the two models can reveal the impact of disclosure of range information on the risky asset market.

### 4.1 The level of asset price

According to the analysis in section 2 and 3, the disclosure of range information $[\underline{v}, \overline{v}]$ would drive asset price from $p_0$ to $p_1$ and the difference between them is

$$p_1 - p_0 = \sigma_\varepsilon \cdot \left[\frac{\psi\left(\frac{\underline{v} - (\tau\tilde{u} + \alpha\tilde{y} + \beta)}{\sigma_\varepsilon}\right) - \psi\left(\frac{\overline{v} - (\tau\tilde{u} + \alpha\tilde{y} + \beta)}{\sigma_\varepsilon}\right)}{\Psi\left(\frac{\overline{v} - (\tau\tilde{u} + \alpha\tilde{y} + \beta)}{\sigma_\varepsilon}\right) - \Psi\left(\frac{\underline{v} - (\tau\tilde{u} + \alpha\tilde{y} + \beta)}{\sigma_\varepsilon}\right)}\right].$$

The midpoint of the disclose range $[\underline{v}, \overline{v}]$ is denoted by $v_m$, that is, $v_m = \frac{\underline{v}+\overline{v}}{2}$. For different ranges with the same length, a range which has a relatively larger midpoint is referred to as a higher range in our discussion. For example, for two ranges $[\underline{v}_1, \overline{v}_1]$ and $[\underline{v}_2, \overline{v}_2]$, if they have the same length and $\frac{\underline{v}_1+\overline{v}_1}{2} > \frac{\underline{v}_2+\overline{v}_2}{2}$, we say that the first range is higher and the second is lower.

For any given range $[\underline{v}, \overline{v}]$, we propose the following lemma to judge whether its disclosure will raise or decrease the price.

**Lemma 1.** If the midpoint of the disclosed range $v_m = \tau\tilde{u} + \alpha\tilde{y} + \beta$, then $p_1 = p_0$. If $v_m < \tau\tilde{u} + \alpha\tilde{y} + \beta$, then $p_1 < p_0$. If $v_m > \tau\tilde{u} + \alpha\tilde{y} + \beta$, then $p_1 > p_0$.

The proof can be seen in appendix A.7. Lemma 1 suggests that whether disclosure of range information can raise asset price depends on the midpoint of the disclosed range. The disclosure of a range with a midpoint higher (lower) than $\tau\tilde{u}+\alpha\tilde{y}+\beta$ can raise (reduce) the price. So, $\tau\tilde{u}+\alpha\tilde{y}+\beta$ can be regarded as a benchmark for assessing the effect of range information disclosure on market price. If we aim to prevent the disclosure of range information from reducing the price, the midpoint of the disclosed range should not be lower than $\tau\tilde{u}+\alpha\tilde{y}+\beta$.

Since

$$\frac{\partial\ \tau\tilde{u}+\alpha\tilde{y}+\beta}{\partial\tilde{u}}=\tau>0\ \ and\ \ \frac{\partial\ \tau\tilde{u}+\alpha\tilde{y}+\beta}{\partial\tilde{y}}=\alpha>0,$$

for preventing the disclosure of range information from reducing asset price, we need to disclose a higher range when the private signal is more optimistic (i.e., $\tilde{u}$ is higher) or the demand of noise trading is larger (i.e., $\tilde{y}$ is higher). In other words, for a certain asset value range $[\underline{v},\bar{v}]$, its disclosure to market is more likely to cause a decrease in asset price when the private signal is more optimistic or the demand of noise trading is larger.

The disclosure of a higher range can boost the confidence of investors about asset value in future and then raise the price. Since the price $p_0$ would be higher when the private signal is more optimistic or the demand of noise trading is larger, the disclosure of a higher range is necessary to maintain the high level of asset price under these conditions.

The disclosed range is regarded as becoming rougher when the lower bound falls and the upper bound rises. By the item (8) in appendix A.2, we have

$$\lim_{\substack{\bar{v}\to+\infty\\ \underline{v}\to-\infty}} p_1=\lim_{\substack{\bar{v}\to+\infty\\ \underline{v}\to-\infty}}\boldsymbol{J}_{[\underline{v},\ \bar{v}]}(\tau\tilde{u}+\alpha\tilde{y}+\beta)=\tau\tilde{u}+\alpha\tilde{y}+\beta=p_0,$$

which suggests the price will approach the level it would be without disclosure of range information when the disclosed range is very rough, that is, the lower bound is very low and the upper bound is very high. In other words, the impact of its disclosure on market price will almost vanish under this case. The intuition is that the disclosure of range information would provide investors with little incremental information when the lower bound is very low and the upper bound is very high, and then can hardly impact market price.

### 4.2 The sensitivity of price to private signal

We analyze the reaction of the market price when the private signal varies, which is regarded as the sensitivity of price to private signal. In the baseline model, the sensitivity of the price to the private signal is

$$u_React_0\triangleq\frac{\partial p_0}{\partial\tilde{u}}=\tau=1-\frac{x_U}{1+\frac{x_I^2\sigma_u^2}{\gamma^2\sigma_\varepsilon^4\sigma_y^2}+\frac{x_I\sigma_u^2}{\sigma_\varepsilon^2}}>0\,. \tag{16}$$

In our main model, by the item (2) in appendix A.2, the sensitivity is

$$u_React_1\triangleq\frac{\partial p_1}{\partial\tilde{u}}=\tau\cdot\boldsymbol{H}_{[\underline{v},\bar{v}]}(\tau\tilde{u}+\alpha\tilde{y}+\beta)>0, \tag{17}$$

where

$$\boldsymbol{H}_{[\underline{v},\overline{v}]}(t) \triangleq \frac{1}{\sigma_\varepsilon^2}\left[\frac{\int_{\underline{v}-t}^{\overline{v}-t} x^2 f_\varepsilon(x)\,dx}{\int_{\underline{v}-t}^{\overline{v}-t} f_\varepsilon(x)\,dx} - \left(\frac{\int_{\underline{v}-t}^{\overline{v}-t} x f_\varepsilon(x)\,dx}{\int_{\underline{v}-t}^{\overline{v}-t} f_\varepsilon(x)\,dx}\right)^2\right]. \tag{18}$$

$u_React_0$ is a positive constant, which means $u_React_0$ is independent of the signal $\tilde{u}$ and noise trading volume $\tilde{y}$ and the price $p_0$ is a linear function of $\tilde{u}$. It's obvious that $u_React_0 < 1$, which implies the variation in the price will be less than that in the signal when the signal varies. In the models of much related literature, the sensitivity of price to signals is also constant and positive.

By equation (17) and the item (2) in appendix A.6, $u_React_1$ is also positive. By the item (3) in appendix A.6 which implies $\boldsymbol{H}_{[\underline{v},\overline{v}]}(\tau\tilde{u} + \alpha\tilde{y} + \beta) < 1$, we have

$$u_React_1 < \tau = u_React_0 < 1.$$

So, our comparison reveals that the disclosure of range information will reduce the sensitivity of price to private signal.

However, according to the item (1) in appendix A.6, $u_React_1$ is not constant and varies continuously with the private signal $\tilde{u}$ and noise trading volume $\tilde{y}$. Figure.1 illustrates the variability and continuity of $u_React_1$, and shows that $u_React_1 < u_React_0$.

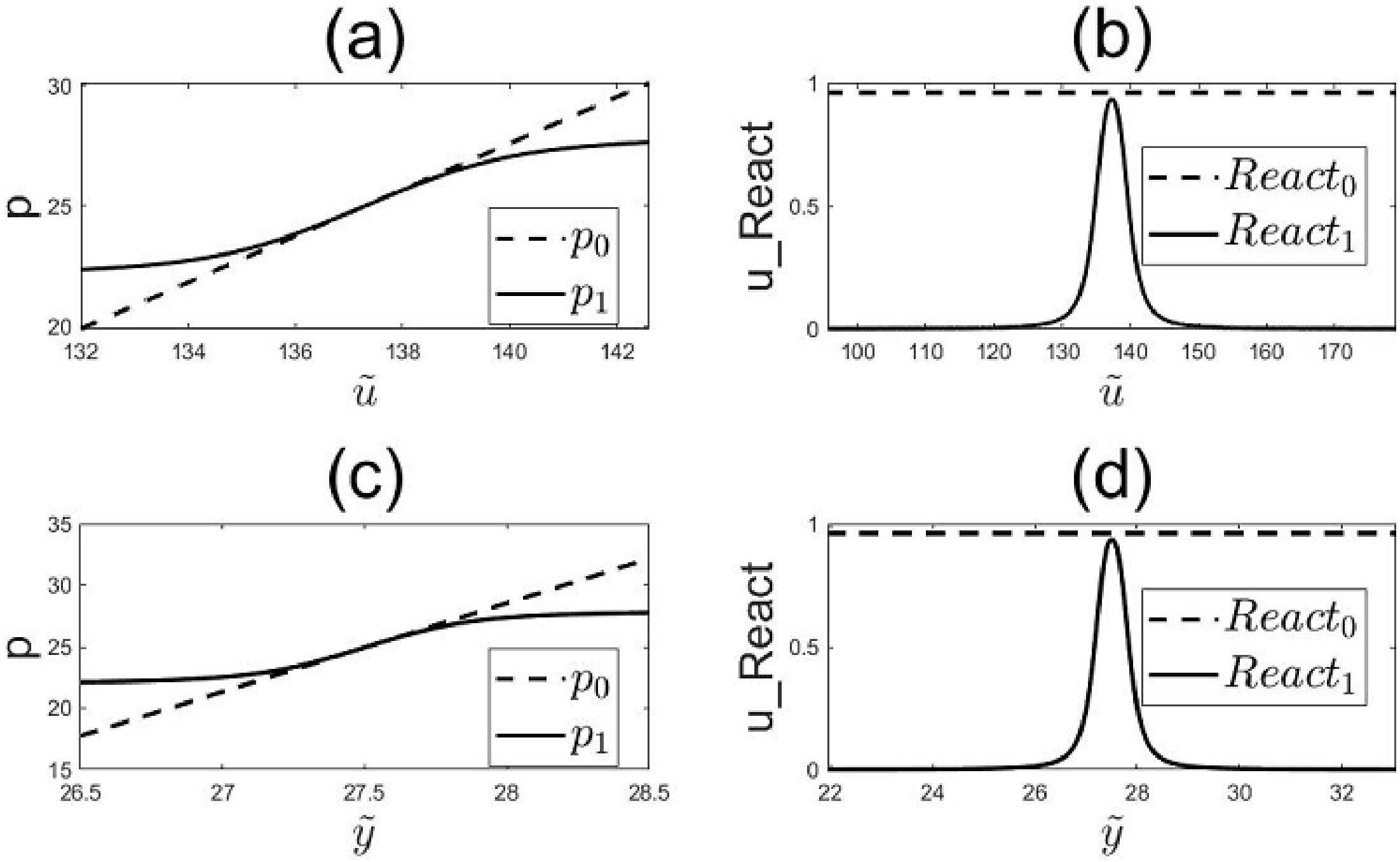

Figure.1. Take the value of $\gamma = 3, \sigma_\varepsilon^2 = 1, \sigma_u^2 = 6, \sigma_y^2 = 5, \mu_0 = 25, x_I = 0.4, Z = 25, \overline{v} = 28, \underline{v} = 22$. In panel (a) and (b), we take $\tilde{y} = 10$, showing the impact of increasing $\tilde{u}$ on price and the sensitivity of price, respectively. In panel (c) and (d), we take $\tilde{u} = 6$, showing the impact of increasing $\tilde{y}$ on price and the sensitivity of price, respectively.

In summary, the comparison implies that disclosure of range information can cause that the reaction of market price to private signal will be no longer constant and vary continuously with private signal and noise trading volume. In the models of related literature, the reaction of price is usually constant or piecewise constant with private signal and noise trading, such as Kodres and Pritsker (2002) [24], Easley et.al. (2014) [7], Mondria et.al. (2022) [20]. So our main model reveals a new pattern of market price reacting to them, which is rarely recorded by existing literature.

Equation (17) shows that the sensitivity of price $p_1$ depends on $[\underline{v}, \overline{v}]$ and $\tau\tilde{u} + \alpha\tilde{y}$, where $\tau\tilde{u} + \alpha\tilde{y}$ is a linear combination of the private signal and noise trading volume.

Denote the length of the range information $[\underline{v}, \overline{v}]$ by $L$, that is, $L \triangleq \overline{v} - \underline{v}$. Now we suppose the length $L$ is fixed. For reflecting the relative position of $[\underline{v}, \overline{v}]$ and the linear combination $\tau\tilde{u} + \alpha\tilde{y}$, we define the distance $d$ between them as

$$d \triangleq \begin{cases} \underline{v} - (\tau\tilde{u} + \alpha\tilde{y}), & if\ \tau\tilde{u} + \alpha\tilde{y} < \underline{v} \\ 0, & if\ \tau\tilde{u} + \alpha\tilde{y} \in [\underline{v}, \overline{v}] \\ (\tau\tilde{u} + \alpha\tilde{y}) - \overline{v}, & if\ \tau\tilde{u} + \alpha\tilde{y} > \overline{v} \end{cases}$$

It is apparent that the sensitivity of price to private signal varies with the distance $d$. We give the following lemma and the proof can be seen in appendix A.8.

**Lemma 2**. $\boldsymbol{H}_{[\underline{v}, \overline{v}]}(\tau\tilde{u} + \alpha\tilde{y} + \beta) > 0$ and $\lim_{d \to +\infty} \boldsymbol{H}_{[\underline{v}, \overline{v}]}(\tau\tilde{u} + \alpha\tilde{y} + \beta) = 0$.

Based on Lemma 2, we have

$$\lim_{d \to +\infty} u_React_1 = \lim_{d \to +\infty} \tau \cdot \boldsymbol{H}_{[\underline{v}, \overline{v}]}(\tau\tilde{u} + \alpha\tilde{y} + \beta) = 0, \tag{19}$$

which implies the sensitivity of price to private signal will approach zero when the distance between $[\underline{v}, \overline{v}]$ and $\tau\tilde{u} + \alpha\tilde{y}$ is very long. In other words, when the linear combination $\tau\tilde{u} + \alpha\tilde{y}$ deviates heavily from the disclosed range $[\underline{v}, \overline{v}]$, market price will hardly react to a variation in private signal. The intuition behind this result is that the reliability of private signal in conveying asset value becomes very low for investors when the linear combination $\tau\tilde{u} + \alpha\tilde{y}$ deviates heavily from the possible range of asset value. Under this case, investors almost no longer rely on the signal to evaluate asset value. So, a variation in the signal can hardly affect investors' trading decisions and then the market price.

Now we suppose the length of range information $L$ is variable. By the item (3) in appendix A.6, we derive

$$\lim_{\substack{\underline{v} \to -\infty \\ \overline{v} \to +\infty}} u_React_1 = \lim_{\substack{\underline{v} \to -\infty \\ \overline{v} \to +\infty}} \tau \cdot \boldsymbol{H}_{[\underline{v}, \overline{v}]}(\tau\tilde{u} + \alpha\tilde{y} + \beta) = \tau = u_React_0.$$

This means the impact of range information disclosure on the sensitivity will approach zero when the disclosed range is very rough. Since $u_React_0 > u_React_1$, the sensitivity of price to private signal has an overall upward trend when the disclosed range becomes increasingly rougher. Intuitively, as the range information becomes rougher, that is, its precision reduces, investors' trading decisions will rely more on the private signal and then the price will be more sensitive to the signal.

### 4.3 The sensitivity of price to range information

We analyze the reaction of market price when the disclosed range varies, which is regarded as the sensitivity of price to range information. First, we suppose the private signal $\tilde{u}$, noise trading $\tilde{y}$ and the lower bound $\underline{v}$ are fixed, and discuss the reaction of market price when the disclosed upper bound $\overline{v}$ varies.

Define the sensitivity of price to the upper bound as

$$\overline{v}_React_1 \triangleq \frac{\partial p_1}{\partial \overline{v}},$$

which can measure the reaction of price when the disclosed upper bound increases by one unit. By the item (7) in appendix A.2,

$$\overline{v}_React_1 = \frac{\partial \boldsymbol{J}_{[\underline{v},\ \overline{v}]}(\tau\tilde{u} + \alpha\tilde{y} + \beta)}{\partial \overline{v}} > 0. \tag{20}$$

So, the price always increases with the upper bound $\overline{v}$, that is, the price will rise if a higher $\overline{v}$ is released to investors. The intuition is that a higher upper bound implies asset value may reach a higher level in future and then investors' expectation of its value will also rise. Denote the distance between the linear combination $\tau\tilde{u} + \alpha\tilde{y}$ and $\overline{v}$ by $d_{\overline{v}}$, that is, $d_{\overline{v}} = |\overline{v} - (\tau\tilde{u} + \alpha\tilde{y})|$. For the scale of the sensitivity $\overline{v}_react_1$, we have the following proposition and its proof is given in appendix A.9.

**Proposition 3.** If $\overline{v}$ is high such that $\overline{v} > \tau\tilde{u} + \alpha\tilde{y}$, then we have

$$\lim_{d_{\overline{v}} \to +\infty} \overline{v}_React_1 = \lim_{d_{\overline{v}} \to +\infty} \frac{\partial \boldsymbol{J}_{[\underline{v},\ \overline{v}]}(\tau\tilde{u} + \alpha\tilde{y} + \beta)}{\partial \overline{v}} = 0\,,$$

which implies a variation in the disclosed upper bound $\overline{v}$ can hardly affect the asset price when $\overline{v}$ far exceeds the linear combination $\tau\tilde{u} + \alpha\tilde{y}$.

Although we have showed raising the released upper bound $\overline{v}$ can boost the asset price, Proposition 3 further suggests that increasingly raising $\overline{v}$ is almost unable to increase the price if $\overline{v}$ has far exceeded the linear combination $\tau\tilde{u} + \alpha\tilde{y}$.

Second, we suppose the private signal $\tilde{u}$, noise trading $\tilde{y}$ and the upper bound $\overline{v}$ are fixed, and discuss the reaction of market price when the disclosed lower bound $\underline{v}$ varies.

Define the sensitivity of price to the lower bound as

$$\underline{v}_React_1 \triangleq \frac{\partial p_1}{\partial \underline{v}},$$

which can measure the reaction of price when the disclosed lower bound increases by one unit. By the item (7) in appendix A.2,

$$\underline{v}_React_1 = \frac{\partial \boldsymbol{J}_{[\underline{v},\ \overline{v}]}(\tau\tilde{u} + \alpha\tilde{y} + \beta)}{\partial \underline{v}} > 0. \tag{21}$$

So, the price will rise if a higher $\underline{v}$ is released to investors. The intuition is that a higher lower bound implies the maximum loss investors may suffer in future declines and then investors' expectation of its value will also rise. Denote the distance between the linear combination $\tau\tilde{u} + \alpha\tilde{y}$ and $\underline{v}$ by $d_{\underline{v}}$, that is, $d_{\underline{v}} = \left|\tau\tilde{u} + \alpha\tilde{y} - \underline{v}\right|$. Then, we have the following proposition and its proof is given in appendix A.10.

**Proposition 4.** If $\underline{v}$ is low such that $\underline{v} < \tau\tilde{u} + \alpha\tilde{y}$, then we have

$$\lim_{d_{\underline{v}} \to +\infty} \underline{v}_React_1 = \lim_{d_{\underline{v}} \to +\infty} \frac{\partial \boldsymbol{J}_{[\underline{v},\ \bar{v}]}(\tau\tilde{u} + \alpha\tilde{y} + \beta)}{\partial \underline{v}} = 0\,,$$

which implies a variation in the disclosed lower bound $\underline{v}$ can hardly affect the asset price when $\underline{v}$ is far below the linear combination $\tau\tilde{u} + \alpha\tilde{y}$.

Although we have showed reducing the released lower bound $\underline{v}$ can decrease the asset price, Proposition 4 further suggests that increasingly reducing $\bar{v}$ is almost unable to decrease the price if $\bar{v}$ has been far below the linear combination $\tau\tilde{u} + \alpha\tilde{y}$.

Third, we suppose the private signal $\tilde{u}$, noise trading $\tilde{y}$ and the length of range information $L$ are fixed, and discuss the reaction of market price when the disclosed range moves. Remind the midpoint of $[\underline{v},\ \bar{v}]$ is denoted by $v_m$.

Since the length of $[\underline{v},\ \bar{v}]$ is fixed, we can use the variation of its midpoints to measure a movement of the range. A one-unit increase (decrease) in the midpoint means a one-unit upward (downward) movement in the disclosed range and vice versa. Thus, we can measure the sensitivity of price to movement in the disclosed range by

$$Range_React_1 \triangleq \frac{\partial p_1}{\partial v_m}.$$

We give the following lemma to specify the sensitivity $Range_react_1$. Its proof is given in appendix A.11.

**Lemma 3.** For any given $\tilde{u}$, $\tilde{y}$ and $L$, we have

$$Range_React_1 = 1 - \boldsymbol{H}_{[\underline{v},\ \bar{v}]}(\tau\tilde{u} + \alpha\tilde{y} + \beta).$$

By the item (2) and (3) in appendix A.6, we have $0 < \boldsymbol{H}_{[\underline{v},\ \bar{v}]}(\tau\tilde{u} + \alpha\tilde{y} + \beta) < 1$ and then

$$1 > Range_React_1 > 0,$$

which means if the disclosed range move upwards, the asset price will rise and the increase in its price will be less than the movement of the disclosed range. In other words, the disclosure of a higher range will cause a higher market price.

Remind that Lemma 2 shows that $\lim_{d \to +\infty} \boldsymbol{H}_{[\underline{v},\bar{v}]}(\tau\tilde{u} + \alpha\tilde{y} + \beta) = 0$. By Lemma 3,

it follows that

$$\lim_{d\to+\infty} Range_React_1 = 1 - \lim_{d\to+\infty} \boldsymbol{H}_{[\underline{v},\overline{v}]}(\tau\tilde{u} + \alpha\tilde{y} + \beta) = 1, \tag{22}$$

which means an upward movement in the disclosed range will cause the asset price to rise by nearly the same amount when $\tau\tilde{u} + \alpha\tilde{y}$ deviates heavily from the disclosed range. The intuition behind this result is that the reliability of private signal about asset value becomes very low for investors when $\tau\tilde{u} + \alpha\tilde{y}$ deviates heavily from the range. Under this case, investors almost wholly rely on the disclosed range to assess future value of asset. Thus, a movement of the disclosed range can efficiently impact investors' assessment about asset value and then the price.

Combining equation (19) with (22), we can propose the following proposition.

**Proposition 5**. If the linear combination $\tau\tilde{u} + \alpha\tilde{y}$ is distant from the disclosed range, a variation in private signal about asset value can hardly affect its price, whereas a movement in the disclosed range can impact the price efficiently.

Furthermore, we can compare the reaction of asset price to variation in the private signal with that to movement in the disclosed range. For any given private signal $\tilde{u}$, noise trading volume $\tilde{y}$ and disclosed range $[\underline{v},\ \overline{v}]$, if $\boldsymbol{H}_{[\underline{v},\overline{v}]}(\tau\tilde{u} + \alpha\tilde{y} + \beta) < \frac{1}{1+\tau}$, then $u_React_1 < Range_React_1$, which means the asset price has a larger reaction to movement in the disclosed range. If $\boldsymbol{H}_{[\underline{v},\overline{v}]}(\tau\tilde{u} + \alpha\tilde{y} + \beta) > \frac{1}{1+\tau}$, then $u_React_1 > Range_React_1$, which means the price has a larger reaction to variation in the private signal. If $\sup_{t\in R} \boldsymbol{H}_{[\underline{v},\overline{v}]}(t) < \frac{1}{1+\tau}$, then $u_React_1 < Range_React_1$ holds for any $\tilde{u}$ and $\tilde{y}$, which implies the reaction of asset price to movement in the disclosed range is always larger than that to variation in the private signal. However, given Proposition 5, the reaction of asset price to variation in the private signal cannot be always larger than that to movement in the disclosed range.

### 4.4 The market liquidity

According to related literature, such as Goldstein and Yang(2017) [22], market liquidity is defined as

$$Liquidity \triangleq \left(\frac{\partial p}{\partial \tilde{y}}\right)^{-1}.$$

A greater market liquidity means noise trading $y$ has a smaller impact on the market price and the market is regarded as deeper and more liquid.

In the equilibrium of the baseline model, the market liquidity is

$$Liquidity_0 = \left(\frac{\partial p_0}{\partial \tilde{y}}\right)^{-1} = \frac{1}{\alpha} = \frac{x_I}{\gamma\sigma_\varepsilon^2}\left(1 - \frac{x_U}{1 + \frac{x_I^2\sigma_u^2}{\gamma^2\sigma_\varepsilon^4\sigma_y^2} + \frac{x_I\sigma_u^2}{\sigma_\varepsilon^2}}\right)^{-1}. \tag{23}$$

In the equilibrium of our main model, by the item (2) in appendix A.2, the market liquidity is

$$Liquidity_1 = \left(\frac{\partial p_1}{\partial \tilde{y}}\right)^{-1} = \frac{1}{\alpha} \cdot \frac{1}{\boldsymbol{H}_{[\underline{v},\overline{v}]}(\tau\tilde{u} + \alpha\tilde{y} + \beta)}. \tag{24}$$

$Liquidity_0$ is constant, that is, it is independent of the private signal $\tilde{u}$ and noise trading $\tilde{y}$. By the item (1) in appendix A.6, $Liquidity_1$ is not constant and varies continuously with $\tilde{u}$ and $\tilde{y}$. Figure.2 illustrates the variability and continuity of $Liquidity_1$.

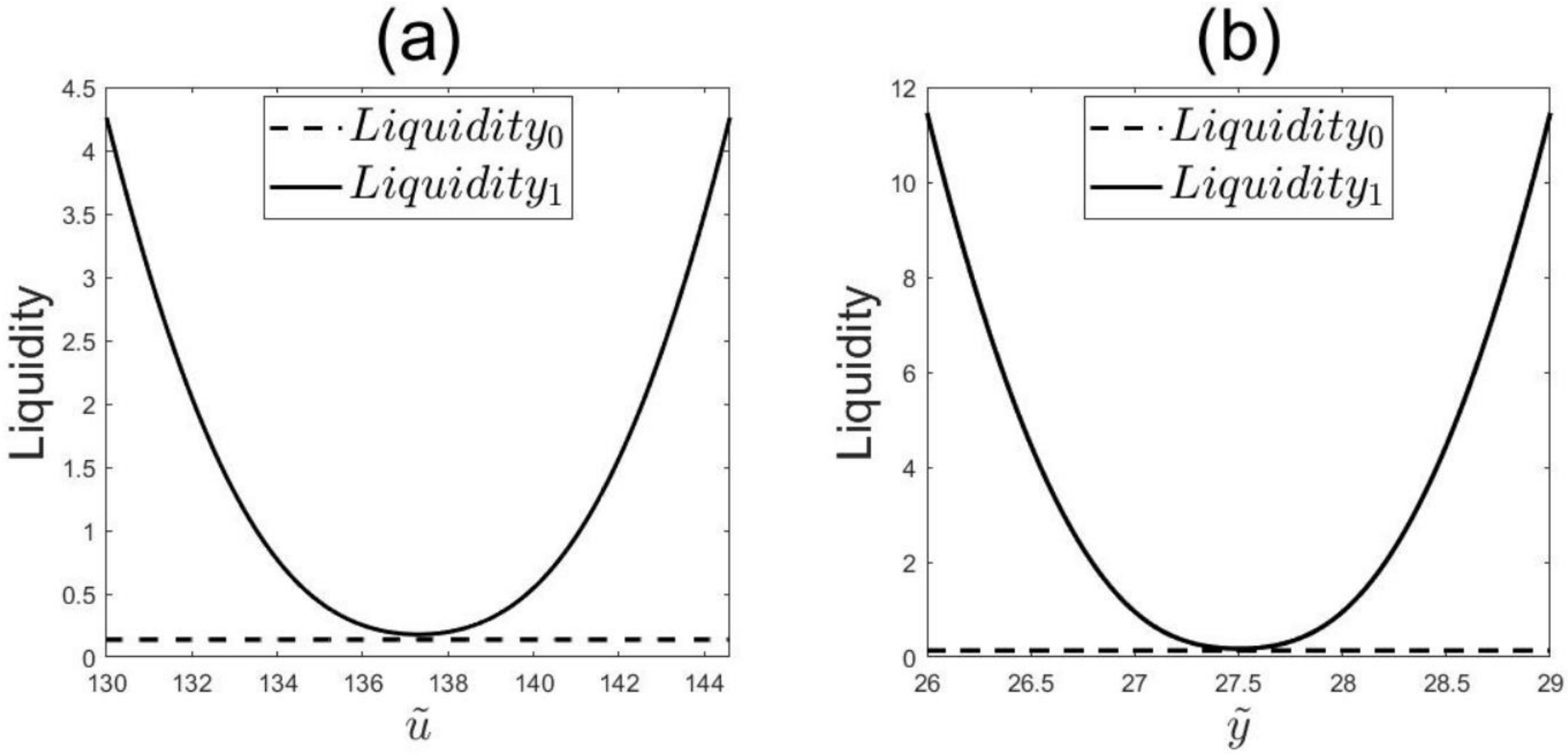

Figure.2. Take the value of $\gamma = 3, \sigma_\varepsilon^2 = 1, \sigma_u^2 = 6, \sigma_y^2 = 5, \mu_0 = 25, x_I = 0.4, Z = 25, \overline{v} = 27, \underline{v} = 23$. In panel (a), we take $\tilde{y} = 10$ and show the impact of increasing $\tilde{u}$ on liquidity. In panel (b), we take $\tilde{u} = 6$ and show the impact of increasing $\tilde{y}$ on liquidity.

According to the item (3) in appendix A.6, we have

$$Liquidity_1 = \frac{1}{\alpha} \cdot \frac{1}{\boldsymbol{H}_{[\underline{v},\overline{v}]}(\tau\tilde{u} + \alpha\tilde{y} + \beta)} > \frac{1}{\alpha} = Liquidity_0$$

holds for any $\tilde{u}$, $\tilde{y}$, $\underline{v}$, $\overline{v}$ with $\underline{v} < \overline{v}$. Thus, it is shown that the disclosure of range information would increase market liquidity, which is illustrated by Figure.2.

In the models of related literature, market liquidity is usually constant or piecewise constant with private signal and noise trading. The comparison between $Liquidity_0$ and $Liquidity_1$ shows that disclosure of range information would cause market liquidity to vary continuously with private signal and noise trading volume. This pattern of liquidity is also rarely recorded by existing related literature.

However, by the item (3) in appendix A.6,

$$\lim_{\substack{\overline{v}\to+\infty \\ \underline{v}\to-\infty}} Liquidity_1 = \frac{1}{\alpha} = Liquidity_0,$$

which means the effect of range information disclosure on market liquidity will approach zero if the disclosed range is very rough. Since $Liquidity_1 > Liquidity_0$, market liquidity has an overall downward trend when the disclosed range information becomes increasingly rougher.

The completeness of investors' information is an important factor that can determine market liquidity. Generally speaking, market liquidity increases with the completeness of investors' information. Disclosure of range information can improve the completeness of investors' information and then raise market liquidity. If the disclosed range is very rough, then it will provide investors with little incremental information and can hardly improve the completeness of investors' information. As a result, the disclosure of a very rough range will generate a minor impact on market liquidity.

Remind that Lemma 2 indicates that $\lim_{d\to+\infty} \boldsymbol{H}_{[\underline{v},\overline{v}]}(\tau\tilde{u}+\alpha\tilde{y}+\beta)=0$. Then, by equation (24), we can derive

$$\lim_{d\to+\infty} Liquidity_1 = \lim_{d\to+\infty} \frac{1}{\alpha}\cdot\frac{1}{\boldsymbol{H}_{[\underline{v},\overline{v}]}(\tau\tilde{u}+\alpha\tilde{y}+\beta)} = +\infty,$$

which implies market will be extremely liquid when the linear combination of private signal and noise trading volume $\tau\tilde{u}+\alpha\tilde{y}$ is far away from the disclosed range.

### 4.5 The asset premium

Asset premium is usually regarded as cost of capital. According to Huang et al.(2020) [21], it is defined as the expected difference between the payoff of asset and its market price. We denote asset premium by $Premium_0$ for the case where range information is not disclosed and by $Premium_1(\underline{v},\overline{v})$ for the case where the range of asset value is disclosed as $[\underline{v},\overline{v}]$. By equation (15) and (10), we have

$$Premium_0 = E[v-p_0] = (1-\tau)\mu_0 - \beta$$

and

$$Premium_1(\underline{v},\overline{v}) = E[v-p_1] = \mu_0 - E\left[\boldsymbol{J}_{[\underline{v},\ \overline{v}]}(\tau u+\alpha y+\beta)\right].$$

In Section 4.2, we have manifested that for any given $\tilde{u}$ and $\tilde{y}$, the price $p_1$ will increase when $\underline{v}$ or $\overline{v}$ are raised. Thus, an increase in the lower bound and upper bound of the disclosed range can both reduce asset premium $Premium_1(\underline{v},\overline{v})$. Similarly, it is also proved in Section 4.2 that an upward movement in the range $[\underline{v},\overline{v}]$ can raise the price $p_1$ for any given $\tilde{u}$ and $\tilde{y}$. Thus, the disclosure of a higher range will also reduce asset premium $Premium_1(\underline{v},\overline{v})$.

For any given range $[\underline{v},\overline{v}]$, we propose the following proposition by which we can judge whether the disclosure of the range will raise or decrease asset premium.

**Proposition 6.** Suppose

$$B_0 \triangleq \beta + \mu_0 \cdot \tau = \mu_0 - Z\gamma\sigma_\varepsilon^2 - \frac{Z\gamma^3 x_U \sigma_u^2 \sigma_\varepsilon^4 \sigma_y^2}{\gamma^2 \sigma_\varepsilon^4 \sigma_y^2 + x_I^2 \sigma_u^2 + \gamma^3 x_I \sigma_u^2 \sigma_\varepsilon^2 \sigma_y^2}. \quad (25)$$

If the midpoint of the disclosed range $v_m = B_0$, then $Premium_1(\underline{v},\overline{v}) = Premium_0$. If $v_m < B_0$, then $Premium_1(\underline{v},\overline{v}) > Premium_0$. If $v_m > B_0$, then $Premium_1(\underline{v},\overline{v}) < Premium_0$.

The proof can be seen in appendix A.12. Proposition 6 suggests that whether disclosure of range information can reduce capital costs depends on the midpoint of the disclosed range. The disclosure of a range with a midpoint higher (lower) than $B_0$ will reduce (raise) capital cost. So, $B_0$ can be regarded as a benchmark for assessing the effect of range information disclosure on capital cost. If we aim to prevent the disclosure of range information from raising capital cost, the midpoint of the disclosed range should be $B_0$ or higher than $B_0$.

By equation (25), we have

$$\frac{\partial B_0}{\partial \sigma_\varepsilon^2} < 0, \frac{\partial B_0}{\partial \sigma_y^2} < 0, \frac{\partial B_0}{\partial \sigma_u^2} < 0, \frac{\partial B_0}{\partial Z} < 0, \frac{\partial B_0}{\partial x_I} > 0, \frac{\partial B_0}{\partial \mu_0} > 0, \quad (26)$$

that is, the benchmark $B_0$ decreases with $\sigma_\varepsilon^2$, $\sigma_y^2$, $\sigma_u^2$, $Z$ and increases with $x_I$, $\mu_0$, which gives rise to the following corollary.

**Corollary 1.** If we aim to prevent the disclosure of range information from raising capital cost, we need to disclose a higher range when

(1) the precision of private signal is higher, that is, $\sigma_\varepsilon^2$ is smaller,

(2) the size of noise trading in market $\sigma_y^2$ is smaller,

(3) the informativeness of private signal $\sigma_u^2$ is smaller,

(4) the supply of risky asset $Z$ is smaller,

(5) the proportion of informed traders $x_I$ is larger,

(6) the unconditional mean of risky asset value $\mu_0$ is higher.

In other words, for a certain asset value range $[\underline{v},\overline{v}]$, its disclosure to market is more likely to cause an increase in asset premium under the six conditions listed in Corollary 1.

Preventing asset premium from rising implies preventing the expectation of asset price from decreasing. In reality, the expectation of asset price is $B_0$ when range information is not disclosed, that is,

$$E[p_0] = \beta + \mu_0 \cdot \tau = B_0.$$

So, the expectation of asset price $E[p_0]$ will be higher under the six conditions listed in Corollary 1. The intuition behind Corollary 1 is that disclosing a higher range is necessary to maintain the high level of $E[p_0]$ under these conditions, that is, to prevent it from decreasing.

We analyze the reaction of asset premium when the disclosed range varies, which is regarded as the sensitivity of asset premium to range information.

The sensitivity of asset premium to the upper and lower bound is respectively

$$\frac{\partial Premium_1(\underline{v}, \overline{v})}{\partial \overline{v}} = -\frac{\partial}{\partial \overline{v}} E\left[\boldsymbol{J}_{[\underline{v},\ \overline{v}]}(\tau u + \alpha y + \beta)\right]$$

and

$$\frac{\partial Premium_1(\underline{v}, \overline{v})}{\partial \underline{v}} = -\frac{\partial}{\partial \underline{v}} E\left[\boldsymbol{J}_{[\underline{v},\ \overline{v}]}(\tau u + \alpha y + \beta)\right].$$

They measure the reaction of asset premium when $\overline{v}$ varies and when $\underline{v}$ varies, respectively.

The distance between $\mu_0$ and $\overline{v}$ and the distance between $\mu_0$ and $\underline{v}$ are denoted by $D_{\overline{v}}$ and $D_{\underline{v}}$, respectively. As for the two sensitivities above, we have the following proposition. Its proof is given in appendix A.13.

**Proposition 7**. If $\overline{v}$ is high such that $\overline{v} > \mu_0$, then we have

$$\lim_{D_{\overline{v}} \to +\infty} \frac{\partial Premium_1(\underline{v}, \overline{v})}{\partial \overline{v}} = 0.$$

Additionally, if $\underline{v}$ is low such that $\underline{v} < \mu_0$, then we have

$$\lim_{D_{\underline{v}} \to +\infty} \frac{\partial Premium_1(\underline{v}, \overline{v})}{\partial \underline{v}} = 0.$$

By Proposition 7, we reveal that a movement in the disclosed upper bound $\overline{v}$ can hardly affect asset premium when $\overline{v}$ far exceeds the unconditional mean of asset value $\mu_0$ and a movement in the disclosed lower bound $\underline{v}$ can hardly affect asset premium when $\underline{v}$ is far below $\mu_0$.

Suppose the length of range information $L$ are fixed, and we focus on the reaction of asset premium when the disclosed range moves and $\mu_0$ varies. For reflecting the relative position of $[\underline{v}, \overline{v}]$ and the unconditional mean of asset value $\mu_0$, we define the distance $D$ between them as

$$D \triangleq \begin{cases} \underline{v} - \mu_0, & if\ \mu_0 < \underline{v} \\ 0, & if\ \mu_0 \in [\underline{v}, \overline{v}] \\ \mu_0 - \overline{v}, & if\ \mu_0 > \overline{v} \end{cases}$$

Remind that the midpoint of $[\underline{v},\ \overline{v}]$ is denoted by $v_m$. Since the length of the range is fixed, its movement can be measured by the variation in its midpoint. Then, the sensitivity of asset premium to movement of the disclosed range can be measured by

$$\frac{\partial Premium_1(\underline{v}, \overline{v})}{\partial v_m} = -\frac{\partial}{\partial v_m} E\left[\boldsymbol{J}_{[\underline{v},\ \overline{v}]}(\tau u + \alpha y + \beta)\right].$$

Meanwhile, the sensitivity of asset premium to variation in $\mu_0$ can be measured by

$$\frac{\partial Premium_1(\underline{v}, \overline{v})}{\partial \mu_0} = -\frac{\partial}{\partial \mu_0} E\left[\boldsymbol{J}_{[\underline{v},\ \overline{v}]}(\tau u + \alpha y + \beta)\right].$$

We have the following proposition, the proof of which is given in appendix A.14.

**Proposition 8**. If the parameters $Z,\ \gamma,\ L,\ \sigma_u^2,\ \sigma_\varepsilon^2,\ \sigma_y^2,\ x_I$ are all fixed, then

$$\lim_{D \to +\infty} \frac{\partial Premium_1(\underline{v}, \overline{v})}{\partial \mu_0} = 0, \qquad \lim_{D \to +\infty} \frac{\partial Premium_1(\underline{v}, \overline{v})}{\partial v_m} = -1.$$

Proposition 8 reveals that if the unconditional mean of asset value $\mu_0$ is distant from the disclosed range, a variation in $\mu_0$ can hardly affect asset premium, whereas a movement of the disclosed range can impact asset premium efficiently. Specifically, an upward movement of the range will cause asset premium to decrease by nearly the same amount when $\mu_0$ is far away from the range.

For the disclosure of rough range information, we have the following proposition about asset premium. The proof is given in appendix A.15.

**Proposition 9**. If the disclosed range is very rough, the effect of range information disclosure on asset premium will almost vanish and its sensitivity to a movement of the disclosed range will almost vanish, that is,

$$\lim_{\substack{\overline{v} \to +\infty \\ \underline{v} \to -\infty}} Premium_1(\underline{v}, \overline{v}) = Premium_0(\underline{v}, \overline{v}),$$

$$\lim_{\substack{\overline{v} \to +\infty \\ \underline{v} \to -\infty}} \frac{\partial Premium_1(\underline{v}, \overline{v})}{\partial v_m} = 0.$$

In summary, if the disclose range is very rough, its disclosure will have a negligible impact on market price and asset premium.

## 5.Conclusion

This study reveals that the disclosure of range information can increase market price or decrease it, which depends on the midpoint of the disclosed range. If the midpoint is higher (lower) than a benchmark which is specified in this study, the disclosure will raise (reduce) the price. The benchmark increases with the demand of noise trading and the private signal received by informed traders.

We also find that the disclosure of range information will decrease the sensitivity of market price to private signal and increase market liquidity. Its disclosure can also change the pattern of market price reacting to private signal and noise trading and the pattern of market liquidity varying with them. The reaction of price to private signal will almost vanish and market liquidity will approach infinity when the linear combination of private signal and noise trading volume deviates heavily from the disclosed range. The sensitivity of price to private signal takes on an overall upward trend and market liquidity takes on an overall downward trend when the range information becomes increasingly rough.

Additionally, the disclosure of range information can increase asset premium or decrease it, which depends on the midpoint of the disclosed range. If the midpoint is higher (lower) than a benchmark which is specified in this study, the disclosure will reduce (raise) asset premium. The benchmark decreases with the size of noise trading, the informativeness of private signal, the supply of risky asset, and increases with the precision of private signal, the proportion of informed traders, and the unconditional mean of risky asset value.

This study further shows that an increase in the upper bound or lower bound of the disclosed range and an upward movement of the range can both raise asset price and then reduce asset premium. However, the impact of a variation in the upper bound on asset price will approach zero when the upper bound far exceeds the linear combination of private signal and noise trading volume, and the impact of a variation in the lower bound on asset price will also approach zero when the lower bound is far below the linear combination of private signal and noise trading volume.

Under the case that the linear combination of private signal and noise trading volume is distant from the disclosed range, a variation in private signal can hardly affect asset price, whereas a movement of the disclosed range can impact the price efficiently. Similarly, when the unconditional mean of asset value is distant from the disclosed range, a variation in the unconditional mean can hardly affect asset premium, whereas a movement of the disclosed range can impact asset premium efficiently.

It is also revealed that range information disclosure will have a negligible impact on market price, liquidity and asset premium if the disclosed range is very rough.

In summary, range information disclosure can significantly affect asset market, but the effect depends on the precision and position of the disclosed range, especially its distance from the linear combination of private signal and noise trading volume, and its distance from the unconditional mean of asset value.

Based on the conclusions, we can give some suggestions about information disclosure for risk management and improvement of market quality. First, when market price is over-sensitive and market liquidity is low, we can reduce the sensitivity of price and increase market liquidity by encouraging more disclosure of range information. Second, for mitigating the risk of market price decline, we should reduce disclosing range information when the private signal about asset value is optimistic or the demand of noise trading is larger. Third, for mitigating the risk of an increase in capital cost, we should reduce disclosing range information when (1) the precision of private signal is higher, (2) the proportion of informed traders is larger, (3) the size of noise trading is smaller, (4) the informativeness of private signal is smaller, (5) the supply of risky asset is smaller. Last, if we attempt to affect asset price by adjusting the information released to market, we should select the appropriate object for adjustment. Adjusting the range information disclosed to market would be more efficient under some cases specified in this study, while adjusting the private signal received by informed traders would be more efficient under other cases.

# Appendix

## A.1 The calculation of equation (2)

Define a conditional distribution $w \triangleq v|(u=\tilde{u})$, then $w \sim N(\tilde{u}, \sigma_\varepsilon^2)$. The distribution and density function of $w$ are denoted by $F_w$ and $f_w$ respectively. The denominator of $U_I(\theta_I)$ is

$$
\begin{aligned}
& E\left[\mathbf{1}_{\{v\in[\underline{v},\overline{v}]\}}|u=\tilde{u}\right] \\
&= E\left[\mathbf{1}_{\{w\in[\underline{v},\overline{v}]\}}\right] \\
&= \int_{\underline{v}}^{\overline{v}} \frac{1}{\sqrt{2\pi}\sigma_\varepsilon} e^{-\frac{(x-\tilde{u})^2}{2\sigma_\varepsilon^2}} dx \\
&= \Psi\left(\frac{\overline{v}-\tilde{u}}{\sigma_\varepsilon}\right) - \Psi\left(\frac{\underline{v}-\tilde{u}}{\sigma_\varepsilon}\right).
\end{aligned}
$$

The numerator of $U_I(\theta_I)$ is

$$
\begin{aligned}
& E\left[-e^{-\gamma(D_0+\theta_I(v-p))}\mathbf{1}_{\{v\in[\underline{v},\overline{v}]\}}|u=\tilde{u}\right] \\
&= E\left[-e^{-\gamma(D_0+\theta_I(w-p))}\mathbf{1}_{\{w\in[\underline{v},\overline{v}]\}}\right] \\
&= -e^{-\gamma D_0+\gamma p\theta_I}\int_{\underline{v}}^{\overline{v}} e^{-\gamma\theta_I x}\cdot\frac{1}{\sqrt{2\pi}\sigma_\varepsilon} e^{-\frac{(x-\tilde{u})^2}{2\sigma_\varepsilon^2}} dx \\
&= -e^{-\gamma D_0+\gamma(p-\tilde{u})\theta_I+\frac{\gamma^2\sigma_\varepsilon^2\theta_I^2}{2}}\left[F_w(\overline{v}+\gamma\sigma_\varepsilon^2\theta_I) - F_w(\underline{v}+\gamma\sigma_\varepsilon^2\theta_I)\right] \\
&= -e^{-\gamma D_0+\gamma(p-\tilde{u})\theta_I+\frac{\gamma^2\sigma_\varepsilon^2\theta_I^2}{2}}\left[\Psi\left(\frac{\overline{v}+\gamma\sigma_\varepsilon^2\theta_I-\tilde{u}}{\sigma_\varepsilon}\right) - \Psi\left(\frac{\underline{v}+\gamma\sigma_\varepsilon^2\theta_I-\tilde{u}}{\sigma_\varepsilon}\right)\right].
\end{aligned}
$$

**A.2 Some properties of the function $\boldsymbol{J}_{[a,b]}(\boldsymbol{t})$**

Remind that $\varepsilon \sim N(0,\sigma_\varepsilon^2)$. Its distribution and density function are denoted by $F_\varepsilon$ and $f_\varepsilon$ respectively. For any fixed $a,\ b \in R$ with $b > a$, the following properties hold.

(1) $a < \boldsymbol{J}_{[a,b]}(t) < b$.

(2) $\boldsymbol{J}_{[a,b]}{}'(t) = \boldsymbol{H}_{[a,b]}(t)$, where

$$\boldsymbol{H}_{[a,b]}(t) \triangleq \frac{1}{\sigma_\varepsilon^2}\left[\frac{\int_{a-t}^{b-t} x^2 f_\varepsilon(x)\,dx}{\int_{a-t}^{b-t} f_\varepsilon(x)\,dx} - \left(\frac{\int_{a-t}^{b-t} x f_\varepsilon(x)\,dx}{\int_{a-t}^{b-t} f_\varepsilon(x)\,dx}\right)^2\right]. \tag{27}$$

(3) $1 > \boldsymbol{J}_{[a,b]}{}'(t) > 0$.

(4) $\lim_{t\to+\infty} \boldsymbol{J}_{[a,b]}(t) = b$ and $\lim_{t\to-\infty} \boldsymbol{J}_{[a,b]}(t) = a$.

(5) $\boldsymbol{J}_{[a,b]}(t)$ is continuous with respect to $t, a, b$.

(6) For any $c \in R$, $\boldsymbol{J}_{[a,b]}(t) = \boldsymbol{J}_{[a-c,b-c]}(t-c) + c$.

(7) $\frac{\partial \boldsymbol{J}_{[a,b]}(t)}{\partial b} = \frac{f_\varepsilon(b-t)\int_{a-t}^{b-t}(b-t-x) f_\varepsilon(x)dx}{\left(\int_{a-t}^{b-t} f_\varepsilon(x)dx\right)^2} > 0$, and

$$\frac{\partial \boldsymbol{J}_{[a,b]}(t)}{\partial a} = \frac{f_\varepsilon(a-t)\int_{a-t}^{b-t}[x-(a-t)] f_\varepsilon(x)dx}{\left(\int_{a-t}^{b-t} f_\varepsilon(x)dx\right)^2} > 0.$$

(8) $\lim_{\substack{a\to-\infty \\ b\to+\infty}} \boldsymbol{J}_{[a,b]}(t) = t$.

**Proof.** We give the proof of the item (1), (3) and (7).

It can be verified that

$$\boldsymbol{J}_{[a,b]}(t) = \frac{\int_{a-t}^{b-t} x f_\varepsilon(x)\,dx}{\int_{a-t}^{b-t} f_\varepsilon(x)\,dx} + t.$$

Then,

$$a = a - t + t < \boldsymbol{J}_{[a,b]}(t) < b - t + t = b.$$

So, the item (1) holds. By some calculations about the derivative of $\boldsymbol{J}_{[a,b]}(\cdot)$, we derive

$$\boldsymbol{J}_{[a,b]}{}'(t) = \frac{\partial}{\partial t}\left[\frac{\int_{a-t}^{b-t} x f_\varepsilon(x)\,dx}{\int_{a-t}^{b-t} f_\varepsilon(x)\,dx} + t\right]$$

$$= \frac{1}{\sigma_\varepsilon^2}\left[\frac{\int_{a-t}^{b-t} x^2 f_\varepsilon(x)\,dx}{\int_{a-t}^{b-t} f_\varepsilon(x)\,dx} - \left(\frac{\int_{a-t}^{b-t} x f_\varepsilon(x)\,dx}{\int_{a-t}^{b-t} f_\varepsilon(x)\,dx}\right)^2\right].$$

Based on the equation above, we have

$$\boldsymbol{J}_{[a,b]}{}'(t) = \frac{1}{\sigma_\varepsilon^2}\left[\frac{\int_{a-t}^{b-t} x^2 f_\varepsilon(x)\,dx}{\int_{a-t}^{b-t} f_\varepsilon(x)\,dx} - \left(\frac{\int_{a-t}^{b-t} x f_\varepsilon(x)\,dx}{\int_{a-t}^{b-t} f_\varepsilon(x)\,dx}\right)^2\right]$$

$$= \frac{\int_{a-t}^{b-t} f_\varepsilon(x)\,dx - (b-t)f_\varepsilon(b-t) + (a-t)f_\varepsilon(a-t)}{\int_{a-t}^{b-t} f_\varepsilon(x)\,dx}$$

$$- \frac{\left(f_\varepsilon(a-t) - f_\varepsilon(b-t)\right)\int_{a-t}^{b-t} x f_\varepsilon(x)\,dx}{\left(\int_{a-t}^{b-t} f_\varepsilon(x)\,dx\right)^2}$$

$$= 1 + \frac{f_\varepsilon(a-t)\int_{a-t}^{b-t}[(a-t)-x]f_\varepsilon(x)\,dx}{\left(\int_{a-t}^{b-t} f_\varepsilon(x)\,dx\right)^2}$$

$$+ \frac{f_\varepsilon(b-t)\int_{a-t}^{b-t}[x-(b-t)]f_\varepsilon(x)\,dx}{\left(\int_{a-t}^{b-t} f_\varepsilon(x)\,dx\right)^2}$$

$$< 1,$$

and

$$\boldsymbol{J}_{[a,\mathrm{b}]}{}'(t) = \frac{1}{\sigma_\varepsilon^2}\left[\frac{\int_{a-t}^{b-t} x^2 f_\varepsilon(x)\,dx}{\int_{a-t}^{b-t} f_\varepsilon(x)\,dx} - \left(\frac{\int_{a-t}^{b-t} x f_\varepsilon(x)\,dx}{\int_{a-t}^{b-t} f_\varepsilon(x)\,dx}\right)^2\right]$$

$$= \frac{\int_{a-t}^{b-t}\left(x - \frac{\int_{a-t}^{b-t} x f_\varepsilon(x)\,dx}{\int_{a-t}^{b-t} f_\varepsilon(x)\,dx}\right)^2 f_\varepsilon(x)\,dx}{\sigma_\varepsilon^2 \cdot \int_{a-t}^{b-t} f_\varepsilon(x)\,dx}$$

$$> 0.$$

So, the item (3) holds.

$$\frac{\partial \boldsymbol{J}_{[a,\mathrm{b}]}(t)}{\partial b} = \frac{\partial}{\partial b}\left[\frac{\int_{a-t}^{b-t} x f_\varepsilon(x)\,dx}{\int_{a-t}^{b-t} f_\varepsilon(x)\,dx} + t\right]$$

$$= \frac{f_\varepsilon(b-t)\int_{a-t}^{b-t}(b-t-x)f_\varepsilon(x)\,dx}{\left(\int_{a-t}^{b-t} f_\varepsilon(x)\,dx\right)^2}$$

$$> 0.$$

$$\frac{\partial \boldsymbol{J}_{[a,\mathrm{b}]}(t)}{\partial a} = \frac{\partial}{\partial a}\left[\frac{\int_{a-t}^{b-t} x f_\varepsilon(x)\,dx}{\int_{a-t}^{b-t} f_\varepsilon(x)\,dx} + t\right]$$

$$= \frac{f_\varepsilon(a-t)\int_{a-t}^{b-t}[x-(a-t)]f_\varepsilon(x)\,dx}{\left(\int_{a-t}^{b-t} f_\varepsilon(x)\,dx\right)^2}$$

$$> 0.$$

Thus, the item (7) is proved.

### A.3 The proof of Proposition 1

It can be proved that there exists a unique $\widetilde{\theta}_I$ such that $U_I{}'(\widetilde{\theta}_I) = 0$ and $U_I(\theta_I)$ reaches the maximum at $\theta_I = \widetilde{\theta}_I$. To define $L(\theta_I) \triangleq \boldsymbol{J}_{[\underline{v},\ \bar{v}]}(\tilde{u} - \gamma\sigma_\varepsilon^2\theta_I) - p$, it follows that $U_I'(\theta_I) = -\gamma \cdot U_I(\theta_I) \cdot L(\theta_I)$. According to the item (3) in appendix A.2, we have

$$L'(\theta_I) = -\gamma\sigma_\varepsilon^2 \cdot \boldsymbol{J}_{[\underline{v},\ \overline{v}]}'(\tilde{u} - \gamma\sigma_\varepsilon^2\theta_I) < 0,$$

that is, $L(\theta_I)$ is a strictly decreasing function for $\theta_I \in R$. Given $\underline{v} < p < \overline{v}$ and the item (4) in appendix A.2, we have

$$\lim_{\theta_I \to -\infty} L(\theta_I) = \overline{v} - p > 0 \ \text{ and } \lim_{\theta_I \to +\infty} L(\theta_I) = \underline{v} - p < 0.$$

In addition, considering the continuity of $L(\theta_I)$, $L(\theta_I) = 0$ definitely has a unique solution $\widetilde{\theta_I}$.

For $\theta_I < \widetilde{\theta_I}$, we have $L(\theta_I) > 0$ and then $U_I'(\theta_I) = -\gamma \cdot U_I(\theta_I) \cdot L(\theta_I) > 0$ (notice that $U_I(\theta_I)$ is always negative). For $\theta_I > \widetilde{\theta_I}$, we have $L(\theta_I) < 0$ and then $U_I'(\theta_I) < 0$. So $\widetilde{\theta_I}$ can maximize the utility $U_I(\theta_I)$. Since $L(\widetilde{\theta_I}) = 0$, we have

$$\boldsymbol{J}_{[\underline{v},\ \overline{v}]}\left(\tilde{u} - \gamma\sigma_\varepsilon^2\widetilde{\theta_I}\right) - p = 0, \tag{28}$$

which means $\widetilde{\theta_I}$ depends on $\tilde{u}$. For a given price $p$, by the implicit function theorem, we have

$$\frac{d\widetilde{\theta_I}}{d\tilde{u}} = -\frac{\dfrac{\partial \boldsymbol{J}_{[\underline{v},\ \overline{v}]}\left(\tilde{u} - \gamma\sigma_\varepsilon^2\widetilde{\theta_I}\right) - p}{\partial\tilde{u}}}{\dfrac{\partial \boldsymbol{J}_{[\underline{v},\ \overline{v}]}\left(\tilde{u} - \gamma\sigma_\varepsilon^2\widetilde{\theta_I}\right) - p}{\partial\widetilde{\theta_I}}}$$

$$= -\frac{\boldsymbol{J}_{[\underline{v},\ \overline{v}]}'\left(\tilde{u} - \gamma\sigma_\varepsilon^2\widetilde{\theta_I}\right)}{\boldsymbol{J}_{[\underline{v},\ \overline{v}]}'\left(\tilde{u} - \gamma\sigma_\varepsilon^2\widetilde{\theta_I}\right) \cdot (-\gamma\sigma_\varepsilon^2)}$$

$$= \frac{1}{\gamma\sigma_\varepsilon^2}.$$

So, we derive a differential equation

$$\frac{d\widetilde{\theta_I}}{d\tilde{u}} = \frac{1}{\gamma\sigma_\varepsilon^2},$$

the solution of which is

$$\widetilde{\theta_I} = \frac{1}{\gamma\sigma_\varepsilon^2} \cdot \tilde{u} + k,$$

where the constant $k$ is independent of $\tilde{u}$. Substituting the above equation of $\widetilde{\theta_I}$ into equation (28), we can derive

$$p = \boldsymbol{J}_{[\underline{v},\ \overline{v}]}(-\gamma\sigma_\varepsilon^2 k),$$

that is, equation (4).

For any fixed $p \in (\underline{v}, \overline{v})$, it can be proved that there exists a unique $k$ satisfying equation (4). By the item (3) in appendix A.2,

$$\frac{\partial \boldsymbol{J}_{[\underline{v},\ \overline{v}]}(-\gamma\sigma_\varepsilon^2 k)}{\partial k}$$

$$= -\gamma\sigma_\varepsilon^2 \cdot \boldsymbol{J}_{[\underline{v},\ \overline{v}]}{}'(-\gamma\sigma_\varepsilon^2 k)$$

$$< 0,$$

which implies that $\boldsymbol{J}_{[\underline{v},\ \overline{v}]}(-\gamma\sigma_\varepsilon^2 k)$ strictly decreases with $k$. By the item (4) in appendix A.2,

$$\lim_{k\to+\infty} \boldsymbol{J}_{[\underline{v},\ \overline{v}]}(-\gamma\sigma_\varepsilon^2 k) = \underline{v}\,, \qquad \lim_{k\to-\infty} \boldsymbol{J}_{[\underline{v},\ \overline{v}]}(-\gamma\sigma_\varepsilon^2 k) = \overline{v}\,.$$

Therefore, there exists a unique $k$ such that $p = \boldsymbol{J}_{[\underline{v},\ \overline{v}]}(-\gamma\sigma_\varepsilon^2 k)$.

By the implicit function theorem,

$$\frac{dk}{dp} = \left[\frac{\partial \boldsymbol{J}_{[\underline{v},\ \overline{v}]}(-\gamma\sigma_\varepsilon^2 k)}{\partial k}\right]^{-1} < 0$$

which implies that $\widetilde{\theta_I} = \frac{1}{\gamma\sigma_\varepsilon^2} \cdot \tilde{u} + k$ strictly decreases with $p$.

### A.4 The calculation of equation (7)

Define a conditional distribution $\eta \triangleq v \left| \left( u + \frac{\alpha}{\tau} y = \frac{J_{[\underline{v},\overline{v}]}^{-1}(p) - \beta}{\tau} \right) \right.$. It follows that $\eta \sim N(\mu_\eta, \sigma_\eta^2)$.

$$U_U(\theta_U; \underline{v}, \overline{v})$$

$$= E\left[ -e^{-\gamma(D_0 + \theta_U(v-p))} \left| u + \frac{\alpha}{\tau} y = \frac{\boldsymbol{J}_{[\underline{v},\overline{v}]}^{-1}(p) - \beta}{\tau}, v \in [\underline{v}, \overline{v}] \right. \right]$$

$$= \frac{E\left[ -e^{-\gamma(D_0 + \theta_U(v-p))} \mathbf{1}_{\{v \in [\underline{v},\overline{v}]\}} \left| u + \frac{\alpha}{\tau} y = \frac{\boldsymbol{J}_{[\underline{v},\overline{v}]}^{-1}(p) - \beta}{\tau} \right. \right]}{E\left[ \mathbf{1}_{\{v \in [\underline{v},\overline{v}]\}} \left| u + \frac{\alpha}{\tau} y = \frac{\boldsymbol{J}_{[\underline{v},\overline{v}]}^{-1}(p) - \beta}{\tau} \right. \right]}$$

$$= \frac{E\left[ -e^{-\gamma(D_0 + \theta_U(\eta - p))} \mathbf{1}_{\{\eta \in [\underline{v},\overline{v}]\}} \right]}{E\left[ \mathbf{1}_{\{\eta \in [\underline{v},\overline{v}]\}} \right]}$$

$$= \frac{-e^{-\gamma D_0 + \gamma p \theta_U} \cdot \int_{\underline{v}}^{\overline{v}} e^{-\gamma\theta_U x} \cdot \frac{1}{\sqrt{2\pi}\sigma_\eta} e^{-\frac{(x-\mu_\eta)^2}{2\sigma_\eta^2}} dx}{\Psi\left(\frac{\overline{v} - \mu_\eta}{\sigma_\eta}\right) - \Psi\left(\frac{\underline{v} - \mu_\eta}{\sigma_\eta}\right)}$$

$$= \frac{-e^{-\gamma D_0+\gamma(p-\mu_\eta)\theta_U+\frac{\gamma^2\sigma_\eta^2\theta_U^2}{2}} \cdot \left[\Psi\left(\frac{\overline{v}+\gamma\sigma_\eta^2\theta_U-\mu_\eta}{\sigma_\eta}\right)-\Psi\left(\frac{\underline{v}+\gamma\sigma_\eta^2\theta_U-\mu_\eta}{\sigma_\eta}\right)\right]}{\Psi\left(\frac{\overline{v}-\mu_\eta}{\sigma_\eta}\right)-\Psi\left(\frac{\underline{v}-\mu_\eta}{\sigma_\eta}\right)}.$$

**A.5 The proof of Proposition 2**

By equation (7),

$$\frac{d\, U_U(\theta_U;\underline{v},\overline{v})}{d\theta_U}$$

$$= -\gamma \cdot U_U(\theta_U;\underline{v},\overline{v}) \cdot \left[\boldsymbol{J}_{[\underline{v},\overline{v}]}(\mu_\eta-\gamma\sigma_\eta^2\theta_U)-p\right].$$

Define $F(\theta_U) \triangleq \boldsymbol{J}_{[\underline{v},\overline{v}]}(\mu_\eta-\gamma\sigma_\eta^2\theta_U)-p$. By the definition of $\boldsymbol{J}_{[\underline{v},\overline{v}]}^{-1}(\cdot)$ and $\overline{\theta}_U$, it can be verified that $F(\overline{\theta}_U)=0$. By the item (3) in appendix A.2, we derive

$$F'(\theta_U) = -\gamma\sigma_\eta^2 \cdot \boldsymbol{J}_{[\underline{v},\overline{v}]}{}'(\mu_\eta-\gamma\sigma_\eta^2\theta_U) < 0,$$

which means $F(\theta_U)$ is a strictly decreasing function of $\theta_U$. So, for any $\theta_U < \overline{\theta}_U$, we have

$$F(\theta_U) > 0 \text{ and } \frac{d\, U_U(\theta_U;\underline{v},\overline{v})}{d\theta_U} = -\gamma \cdot U_U(\theta_U;\underline{v},\overline{v}) \cdot F(\theta_U) > 0.$$

For any $\theta_U > \overline{\theta}_U$, we have

$$F(\theta_U) < 0 \text{ and } \frac{d\, U_U(\theta_U;\underline{v},\overline{v})}{d\theta_U} = -\gamma \cdot U_U(\theta_U;\underline{v},\overline{v}) \cdot F(\theta_U) < 0.$$

(Notice that $U_U(\cdot\,;\underline{v},\overline{v}) < 0$.)

It follows that $U_U(\theta_U;\underline{v},\overline{v})$ reaches its maximum at $\theta_U = \overline{\theta}_U$.

**A.6 Some properties of the function $\boldsymbol{H}_{[a,b]}(t)$**

For any fixed $a,\ b \in R$ with $b > a$, the following properties hold.

(1) $\boldsymbol{H}_{[a,b]}(t)$ is continuous with respect to $t,\ a,\ b$.

(2) For any $t \in R$, $\boldsymbol{H}_{[a,b]}(t) > 0$.

(3) For any $t \in R$, $\boldsymbol{H}_{[a,b]}(t) < 1$ and $\lim_{\substack{a\to-\infty \\ b\to+\infty}} \boldsymbol{H}_{[a,b]}(t) = 1$.

(4) $\lim_{t\to+\infty} \boldsymbol{H}_{[a,b]}(t) = 0$ and $\lim_{t\to-\infty} \boldsymbol{H}_{[a,b]}(t) = 0$.

(5) For any $c \in R$, $\boldsymbol{H}_{[a,b]}(t) = \boldsymbol{H}_{[a-c,b-c]}(t-c)$.

**Proof.** We give the proof of the item (4). In appendix A.2, we have showed $\boldsymbol{H}_{[a,b]}(t) = \boldsymbol{J}_{[a,b]}{}'(t) < 1$ and $\boldsymbol{H}_{[a,b]}(t) > 0$.

When $t \to -\infty$, we have

$$1 \leftarrow \frac{a-t}{b-t} < \frac{a-t}{\frac{\int_{a-t}^{b-t} x f_\varepsilon(x)\,dx}{\int_{a-t}^{b-t} f_\varepsilon(x)\,dx}} < \frac{a-t}{a-t} = 1,$$

which implies

$$\lim_{t\to-\infty} \frac{(a-t)\cdot\int_{a-t}^{b-t} f_\varepsilon(x)\,dx}{\int_{a-t}^{b-t} x f_\varepsilon(x)\,dx} = 1.$$

By the definition of $\boldsymbol{H}_{[a,b]}(t)$ (i.e., equation (27)), we have

$$\lim_{t\to-\infty} \boldsymbol{H}_{[a,b]}(t) = \frac{1}{\sigma_\varepsilon^2}\cdot\lim_{t\to-\infty}\left[\frac{\int_{a-t}^{b-t} x^2 f_\varepsilon(x)\,dx}{\int_{a-t}^{b-t} f_\varepsilon(x)\,dx} - (a-t)^2\right] - \frac{1}{\sigma_\varepsilon^2}\cdot\lim_{t\to-\infty}\left[\left(\frac{\int_{a-t}^{b-t} x f_\varepsilon(x)\,dx}{\int_{a-t}^{b-t} f_\varepsilon(x)\,dx}\right)^2 - (a-t)^2\right]. \tag{29}$$

The first part is

$$\lim_{t\to-\infty}\left[\frac{\int_{a-t}^{b-t} x^2 f_\varepsilon(x)\,dx}{\int_{a-t}^{b-t} f_\varepsilon(x)\,dx} - (a-t)^2\right]$$

$$= \lim_{t\to-\infty} \frac{\int_{a-t}^{b-t} x^2 f_\varepsilon(x)\,dx - (a-t)^2\cdot\int_{a-t}^{b-t} f_\varepsilon(x)\,dx}{\int_{a-t}^{b-t} f_\varepsilon(x)\,dx}$$

$$= \lim_{t\to-\infty}\left\{\frac{(a-t)^2 f_\varepsilon(a-t) - (b-t)^2 f_\varepsilon(b-t)}{f_\varepsilon(a-t) - f_\varepsilon(b-t)} + \frac{2(a-t)\int_{a-t}^{b-t} f_\varepsilon(x)\,dx - (a-t)^2\cdot[f_\varepsilon(a-t) - f_\varepsilon(b-t)]}{f_\varepsilon(a-t) - f_\varepsilon(b-t)}\right\}$$

$$= \lim_{t\to-\infty} \frac{(a-b)(a+b-2t) f_\varepsilon(b-t)}{f_\varepsilon(a-t) - f_\varepsilon(b-t)} + \lim_{t\to-\infty} \frac{2(a-t)\int_{a-t}^{b-t} f_\varepsilon(x)\,dx}{f_\varepsilon(a-t) - f_\varepsilon(b-t)}$$

$$= \lim_{t\to-\infty} \frac{(a-b)(a+b-2t)}{e^{-\frac{(b-a)(2t-a-b)}{2\sigma_\varepsilon^2}} - 1} + \lim_{t\to-\infty} \frac{2\sigma_\varepsilon^2 (a-t)\int_{a-t}^{b-t} f_\varepsilon(x)\,dx}{\int_{a-t}^{b-t} x f_\varepsilon(x)\,dx}$$

$$= 2\sigma_\varepsilon^2.$$

The second part is

$$\lim_{t\to-\infty}\left[\left(\frac{\int_{a-t}^{b-t} x f_\varepsilon(x)\,dx}{\int_{a-t}^{b-t} f_\varepsilon(x)\,dx}\right)^2 - (a-t)^2\right]$$

$$= \lim_{t\to-\infty} \frac{\left(\int_{a-t}^{b-t} x f_\varepsilon(x)\,dx\right)^2 - (a-t)^2\cdot\left(\int_{a-t}^{b-t} f_\varepsilon(x)\,dx\right)^2}{\left(\int_{a-t}^{b-t} f_\varepsilon(x)\,dx\right)^2}$$

$$= \lim_{t\to-\infty} \left\{ \frac{2\int_{a-t}^{b-t} x f_\varepsilon(x)\,dx[(a-t)f_\varepsilon(a-t)-(b-t)f_\varepsilon(b-t)]}{2\int_{a-t}^{b-t} f_\varepsilon(x)\,dx\cdot[f_\varepsilon(a-t)-f_\varepsilon(b-t)]} \right.$$

$$\left. + \frac{2(a-t)\left(\int_{a-t}^{b-t} f_\varepsilon(x)\,dx\right)^2 - 2(a-t)^2[f_\varepsilon(a-t)-f_\varepsilon(b-t)]\int_{a-t}^{b-t} f_\varepsilon(x)\,dx}{2\int_{a-t}^{b-t} f_\varepsilon(x)\,dx\cdot[f_\varepsilon(a-t)-f_\varepsilon(b-t)]} \right\}$$

$$= \lim_{t\to-\infty} \frac{f_\varepsilon(b-t)\cdot\int_{a-t}^{b-t}[(a-t)^2-(b-t)x]f_\varepsilon(x)\,dx}{\int_{a-t}^{b-t} f_\varepsilon(x)\,dx\cdot[f_\varepsilon(a-t)-f_\varepsilon(b-t)]}$$

$$+ \lim_{t\to-\infty} \frac{f_\varepsilon(a-t)\cdot\int_{a-t}^{b-t}[(a-t)x-(a-t)^2]f_\varepsilon(x)\,dx}{\int_{a-t}^{b-t} f_\varepsilon(x)\,dx\cdot[f_\varepsilon(a-t)-f_\varepsilon(b-t)]}$$

$$+ \lim_{t\to-\infty} \frac{(a-t)\int_{a-t}^{b-t} f_\varepsilon(x)\,dx}{[f_\varepsilon(a-t)-f_\varepsilon(b-t)]}, \tag{30}$$

When $t \to -\infty$, we have

$$0 < \left| \frac{f_\varepsilon(b-t)\cdot\int_{a-t}^{b-t}[(a-t)^2-(b-t)x]f_\varepsilon(x)\,dx}{\int_{a-t}^{b-t} f_\varepsilon(x)\,dx\cdot[f_\varepsilon(a-t)-f_\varepsilon(b-t)]} \right|$$

$$< \frac{f_\varepsilon(b-t)\cdot[(b-t)^2-(a-t)^2]}{f_\varepsilon(a-t)-f_\varepsilon(b-t)} = \frac{(b-a)(b+a-2t)}{e^{-\frac{(b-a)(2t-a-b)}{2\sigma_\varepsilon^2}}-1} \to 0,$$

which implies

$$\lim_{t\to-\infty} \frac{f_\varepsilon(b-t)\cdot\int_{a-t}^{b-t}[(a-t)^2-(b-t)x]f_\varepsilon(x)\,dx}{\int_{a-t}^{b-t} f_\varepsilon(x)\,dx\cdot[f_\varepsilon(a-t)-f_\varepsilon(b-t)]} = 0.$$

$$\lim_{t\to-\infty} \frac{f_\varepsilon(a-t)\cdot\int_{a-t}^{b-t}[(a-t)x-(a-t)^2]f_\varepsilon(x)\,dx}{\int_{a-t}^{b-t} f_\varepsilon(x)\,dx\cdot[f_\varepsilon(a-t)-f_\varepsilon(b-t)]}$$

$$= \lim_{t\to-\infty} \frac{\int_{a-t}^{b-t}[(a-t)x-(a-t)^2]f_\varepsilon(x)\,dx}{\int_{a-t}^{b-t} f_\varepsilon(x)\,dx} \cdot \lim_{t\to-\infty} \frac{f_\varepsilon(a-t)}{f_\varepsilon(a-t)-f_\varepsilon(b-t)}$$

$$= \lim_{t\to-\infty} \left\{ \frac{(a-t)[(a-t)f_\varepsilon(a-t)-(b-t)f_\varepsilon(b-t)] - \int_{a-t}^{b-t} x f_\varepsilon(x)\,dx}{f_\varepsilon(a-t)-f_\varepsilon(b-t)} \right.$$

$$\left. + \frac{2(a-t)\int_{a-t}^{b-t} f_\varepsilon(x)\,dx - (a-t)^2[f_\varepsilon(a-t)-f_\varepsilon(b-t)]}{f_\varepsilon(a-t)-f_\varepsilon(b-t)} \right\}$$

$$\cdot \lim_{t\to-\infty} \frac{e^{-\frac{(b-a)(2t-a-b)}{2\sigma_\varepsilon^2}}}{e^{-\frac{(b-a)(2t-a-b)}{2\sigma_\varepsilon^2}}-1}$$

$$= \lim_{t\to-\infty} \frac{\int_{a-t}^{b-t} x f_\varepsilon(x)\,dx}{f_\varepsilon(b-t) - f_\varepsilon(a-t)} + \lim_{t\to-\infty} \frac{(b-a)(a-t) f_\varepsilon(b-t)}{f_\varepsilon(b-t) - f_\varepsilon(a-t)}$$

$$+ \lim_{t\to-\infty} \frac{2(a-t)\int_{a-t}^{b-t} f_\varepsilon(x)\,dx}{f_\varepsilon(a-t) - f_\varepsilon(b-t)}$$

$$= -\sigma_\varepsilon^2 + \lim_{t\to-\infty} \frac{(b-a)(a-t)}{1 - e^{-\frac{(b-a)(2t-a-b)}{2\sigma_\varepsilon^2}}} + \lim_{t\to-\infty} \frac{2\sigma_\varepsilon^2 (a-t)\int_{a-t}^{b-t} f_\varepsilon(x)\,dx}{\int_{a-t}^{b-t} x f_\varepsilon(x)\,dx}$$

$$= \sigma_\varepsilon^2.$$

$$\lim_{t\to-\infty} \frac{(a-t)\int_{a-t}^{b-t} f_\varepsilon(x)\,dx}{[f_\varepsilon(a-t) - f_\varepsilon(b-t)]} = \lim_{t\to-\infty} \frac{\sigma_\varepsilon^2 (a-t)\int_{a-t}^{b-t} f_\varepsilon(x)\,dx}{\int_{a-t}^{b-t} x f_\varepsilon(x)\,dx} = \sigma_\varepsilon^2.$$

By equation (30), we derive

$$\lim_{t\to-\infty} \left[ \left( \frac{\int_{a-t}^{b-t} x f_\varepsilon(x)\,dx}{\int_{a-t}^{b-t} f_\varepsilon(x)\,dx} \right)^2 - (a-t)^2 \right] = 0 + \sigma_\varepsilon^2 + \sigma_\varepsilon^2 = 2\sigma_\varepsilon^2.$$

By equation (29), it follows that

$$\lim_{t\to-\infty} \boldsymbol{H}_{[a,b]}(t) = \frac{1}{\sigma_\varepsilon^2} \cdot (2\sigma_\varepsilon^2) - \frac{1}{\sigma_\varepsilon^2} \cdot (2\sigma_\varepsilon^2) = 0.$$

According to the definition of $\boldsymbol{H}_{[a,b]}(t)$ (i.e., equation (27)), it is easy to show

$$\boldsymbol{H}_{[a,b]}(t) = \boldsymbol{H}_{[-b,-a]}(-t).$$

Then,

$$\lim_{t\to+\infty} \boldsymbol{H}_{[a,b]}(t) = \lim_{t\to+\infty} \boldsymbol{H}_{[-b,-a]}(-t) = \lim_{s\to-\infty} \boldsymbol{H}_{[-b,-a]}(s) = 0.$$

So, we have showed the item (4) holds.

**A.7 The proof of Lemma 1**

Remind $\psi$ is the density function of the normal standard distribution.

$$p_1 - p_0 = \sigma_\varepsilon \cdot \left[ \frac{\psi\left(\frac{\underline{v} - (\tau\tilde{u} + \alpha\tilde{y} + \beta)}{\sigma_\varepsilon}\right) - \psi\left(\frac{\overline{v} - (\tau\tilde{u} + \alpha\tilde{y} + \beta)}{\sigma_\varepsilon}\right)}{\Psi\left(\frac{\overline{v} - (\tau\tilde{u} + \alpha\tilde{y} + \beta)}{\sigma_\varepsilon}\right) - \Psi\left(\frac{\underline{v} - (\tau\tilde{u} + \alpha\tilde{y} + \beta)}{\sigma_\varepsilon}\right)} \right].$$

If $v_m = \tau\tilde{u} + \alpha\tilde{y} + \beta$, $\left|\frac{\underline{v}-(\tau\tilde{u}+\alpha\tilde{y}+\beta)}{\sigma_\varepsilon}\right| = \left|\frac{\overline{v}-(\tau\tilde{u}+\alpha\tilde{y}+\beta)}{\sigma_\varepsilon}\right|$ and then $p_1 = p_0$.

If $v_m < \tau\tilde{u} + \alpha\tilde{y} + \beta$, $\left|\frac{\underline{v}-(\tau\tilde{u}+\alpha\tilde{y}+\beta)}{\sigma_\varepsilon}\right| > \left|\frac{\overline{v}-(\tau\tilde{u}+\alpha\tilde{y}+\beta)}{\sigma_\varepsilon}\right|$ and then $p_1 < p_0$.

If $v_m > \tau\tilde{u} + \alpha\tilde{y} + \beta$, $\left|\frac{\underline{v}-(\tau\tilde{u}+\alpha\tilde{y}+\beta)}{\sigma_\varepsilon}\right| < \left|\frac{\overline{v}-(\tau\tilde{u}+\alpha\tilde{y}+\beta)}{\sigma_\varepsilon}\right|$ and then $p_1 > p_0$.

**A.8 The proof of Lemma 2**

The item (2) in appendix A.6 directly suggests

$$\boldsymbol{H}_{[\underline{v},\overline{v}]}(\tau\tilde{u} + \alpha\tilde{y} + \beta) > 0$$

By the item (5) in appendix A.6, we can derive

$$\boldsymbol{H}_{[\underline{v},\overline{v}]}(\tau\tilde{u}+\alpha\tilde{y}+\beta)=\boldsymbol{H}_{[0,\overline{v}-\underline{v}]}\left(\tau\tilde{u}+\alpha\tilde{y}-\underline{v}+\beta\right)$$
$$=\boldsymbol{H}_{[0,L]}(\beta-d)$$

if $\tau\tilde{u}+\alpha\tilde{y}<\underline{v}$. If $\tau\tilde{u}+\alpha\tilde{y}>\overline{v}$, we can derive

$$\boldsymbol{H}_{[\underline{v},\overline{v}]}(\tau\tilde{u}+\alpha\tilde{y}+\beta)=\boldsymbol{H}_{[\underline{v}-\overline{v},0]}(\tau\tilde{u}+\alpha\tilde{y}-\overline{v}+\beta)$$
$$=\boldsymbol{H}_{[-L,0]}(\beta+d).$$

By the item (4) in appendix A.6, we have

$$\lim_{d\to+\infty}\boldsymbol{H}_{[0,L]}(\beta-d)=0 \text{ and } \lim_{d\to+\infty}\boldsymbol{H}_{[-L,0]}(\beta+d)=0.$$

It follows that

$$\lim_{d\to+\infty}\boldsymbol{H}_{[\underline{v},\overline{v}]}(\tau\tilde{u}+\alpha\tilde{y}+\beta)=0.$$

**A.9 The proof of Proposition 3**

Remind that $d_{\overline{v}}=\overline{v}-(\tau\tilde{u}+\alpha\tilde{y})$. Then, By the item (7) in appendix A.2, we have

$$\overline{v}_React_1=\frac{\partial\boldsymbol{J}_{[\underline{v},\ \overline{v}]}(\tau\tilde{u}+\alpha\tilde{y}+\beta)}{\partial\overline{v}}$$
$$=\frac{f_\varepsilon(d_{\overline{v}}-\beta)\int_{\underline{v}-(\tau\tilde{u}+\alpha\tilde{y}+\beta)}^{d_{\overline{v}}-\beta}[d_{\overline{v}}-\beta-x]f_\varepsilon(x)\,dx}{\left(\int_{\underline{v}-(\tau\tilde{u}+\alpha\tilde{y}+\beta)}^{d_{\overline{v}}-\beta}f_\varepsilon(x)\,dx\right)^2}$$

The numerator of $\overline{v}_React_1$ satisfies

$$0<f_\varepsilon(d_{\overline{v}}-\beta)\int_{\underline{v}-(\tau\tilde{u}+\alpha\tilde{y}+\beta)}^{d_{\overline{v}}-\beta}[d_{\overline{v}}-\beta-x]f_\varepsilon(x)\,dx$$
$$\leq\frac{\left[d_{\overline{v}}-\left(\underline{v}-(\tau\tilde{u}+\alpha\tilde{y})\right)\right]^2}{2\pi\sigma_\varepsilon^2}\cdot e^{-\frac{(d_{\overline{v}}-\beta)^2}{2\sigma_\varepsilon^2}}\xrightarrow{when\ d_{\overline{v}}\to+\infty}0.$$

The denominator of $\underline{v}_React_1$ satisfies

$$\lim_{d_{\overline{v}}\to+\infty}\left(\int_{\underline{v}-(\tau\tilde{u}+\alpha\tilde{y}+\beta)}^{d_{\overline{v}}-\beta}f_\varepsilon(x)\,dx\right)^2=\left(\int_{\underline{v}-(\tau\tilde{u}+\alpha\tilde{y}+\beta)}^{+\infty}f_\varepsilon(x)\,dx\right)^2>0.$$

Thus,

$$\lim_{d_{\overline{v}}\to+\infty}\overline{v}_React_1=0.$$

**A.10 The proof of Proposition 4**

Remind $d_{\underline{v}}=(\tau\tilde{u}+\alpha\tilde{y})-\underline{v}$. Then, by the item (7) in appendix A.2, we have

$$\underline{v}_React_1=\frac{\partial\boldsymbol{J}_{[\underline{v},\ \overline{v}]}(\tau\tilde{u}+\alpha\tilde{y}+\beta)}{\partial\underline{v}}$$

$$
= \frac{f_\varepsilon\left(-d_{\underline{v}}-\beta\right)\int_{-d_{\underline{v}}-\beta}^{\overline{v}-(\tau\tilde{u}+\alpha\tilde{y}+\beta)}\left[x+\beta+d_{\underline{v}}\right]f_\varepsilon(x)\,dx}{\left(\int_{-d_{\underline{v}}-\beta}^{\overline{v}-(\tau\tilde{u}+\alpha\tilde{y}+\beta)}f_\varepsilon(x)\,dx\right)^2}.
$$

The numerator of $\underline{v}_React_1$ satisfies

$$
0 < f_\varepsilon\left(-d_{\underline{v}}-\beta\right)\int_{-d_{\underline{v}}-\beta}^{\overline{v}-(\tau\tilde{u}+\alpha\tilde{y}+\beta)}\left[x+\beta+d_{\underline{v}}\right]f_\varepsilon(x)\,dx
$$

$$
\leq \frac{[\overline{v}-(\tau\tilde{u}+\alpha\tilde{y})+d_{\overline{v}}]^2}{2\pi\sigma_\varepsilon^2}\cdot e^{-\frac{\left(-d_{\underline{v}}-\beta\right)^2}{2\sigma_\varepsilon^2}} \xrightarrow{when\ d_{\underline{v}}\to+\infty} 0.
$$

The denominator of $\underline{v}_React_1$ satisfies

$$
\lim_{d_{\underline{v}}\to+\infty}\left(\int_{-d_{\underline{v}}-\beta}^{\overline{v}-(\tau\tilde{u}+\alpha\tilde{y}+\beta)}f_\varepsilon(x)\,dx\right)^2 = \left(\int_{-\infty}^{\overline{v}-(\tau\tilde{u}+\alpha\tilde{y}+\beta)}f_\varepsilon(x)\,dx\right)^2 > 0
$$

Thus,

$$
\lim_{d_{\underline{v}}\to+\infty}\underline{v}_React_1 = 0.
$$

### A.11 The proof of Lemma 3

$[\,\underline{v},\ \overline{v}\,]$ can be expressed as $\left[v_m-\frac{L}{2}\,,\ v_m+\frac{L}{2}\right]$. By the item (6) in appendix A.2,

$$
\boldsymbol{J}_{\left[v_m-\frac{L}{2},v_m+\frac{L}{2}\right]}(\tau\tilde{u}+\alpha\tilde{y}+\beta)
$$

$$
= \boldsymbol{J}_{\left[-\frac{L}{2},\ \frac{L}{2}\right]}(\tau\tilde{u}+\alpha\tilde{y}+\beta-v_m)+v_m.
$$

Since the length of the range information $L$ is fixed, we have

$$
\begin{aligned}
Range_React_1 &= \frac{\partial\boldsymbol{J}_{[\underline{v},\ \overline{v}]}(\tau\tilde{u}+\alpha\tilde{y}+\beta)}{\partial v_m} \\
&= \frac{\partial\boldsymbol{J}_{\left[v_m-\frac{L}{2},\ v_m+\frac{L}{2}\right]}(\tau\tilde{u}+\alpha\tilde{y}+\beta)}{\partial v_m} \\
&= \frac{\partial}{\partial v_m}\left[\boldsymbol{J}_{\left[-\frac{L}{2},\ \frac{L}{2}\right]}(\tau\tilde{u}+\alpha\tilde{y}+\beta-v_m)+v_m\right] \\
&= 1-\boldsymbol{H}_{\left[-\frac{L}{2},\ \frac{L}{2}\right]}(\tau\tilde{u}+\alpha\tilde{y}+\beta-v_m) \\
&= 1-\boldsymbol{H}_{\left[-\frac{L}{2}+v_m,\ \frac{L}{2}+v_m\right]}(\tau\tilde{u}+\alpha\tilde{y}+\beta) \\
&= 1-\boldsymbol{H}_{[\underline{v},\ \overline{v}]}(\tau\tilde{u}+\alpha\tilde{y}+\beta),
\end{aligned}
$$

where the fourth and fifth equality follow from the item (2) in appendix A.2 and the item (5) in appendix A.6 respectively.

**A.12 The proof of Proposition 6**

Define $X \triangleq \tau u + \alpha y + \beta$, then $X \sim N(B_0, \sigma_X^2)$ with $\sigma_X^2 = \tau^2 \sigma_u^2 + \alpha^2 \sigma_y^2$. Define

$$T(x; \underline{v}, \overline{v}) \triangleq \frac{\psi\left(\frac{\underline{v} - x}{\sigma_\varepsilon}\right) - \psi\left(\frac{\overline{v} - x}{\sigma_\varepsilon}\right)}{\Psi\left(\frac{\overline{v} - x}{\sigma_\varepsilon}\right) - \Psi\left(\frac{\underline{v} - x}{\sigma_\varepsilon}\right)}.$$

First of all, we can prove that $T(-x + B_0; \underline{v}, \overline{v}) = -T(x + B_0; \underline{v}, \overline{v})$ if the midpoint of the disclosed range is $B_0$ (i.e., $\overline{v} - B_0 = -(\underline{v} - B_0)$).

$$T(-x + B_0; \underline{v}, \overline{v}) = \frac{\psi\left(\frac{\underline{v} - B_0}{\sigma_\varepsilon} + \frac{x}{\sigma_\varepsilon}\right) - \psi\left(\frac{\overline{v} - B_0}{\sigma_\varepsilon} + \frac{x}{\sigma_\varepsilon}\right)}{\Psi\left(\frac{\overline{v} - B_0}{\sigma_\varepsilon} + \frac{x}{\sigma_\varepsilon}\right) - \Psi\left(\frac{\underline{v} - B_0}{\sigma_\varepsilon} + \frac{x}{\sigma_\varepsilon}\right)}$$

$$= \frac{\psi\left(-\frac{\underline{v} - B_0}{\sigma_\varepsilon} - \frac{x}{\sigma_\varepsilon}\right) - \psi\left(-\frac{\overline{v} - B_0}{\sigma_\varepsilon} - \frac{x}{\sigma_\varepsilon}\right)}{\left[1 - \Psi\left(-\frac{\overline{v} - B_0}{\sigma_\varepsilon} - \frac{x}{\sigma_\varepsilon}\right)\right] - \left[1 - \Psi\left(-\frac{\underline{v} - B_0}{\sigma_\varepsilon} - \frac{x}{\sigma_\varepsilon}\right)\right]}$$

$$= \frac{\psi\left(\frac{\overline{v} - B_0}{\sigma_\varepsilon} - \frac{x}{\sigma_\varepsilon}\right) - \psi\left(\frac{\underline{v} - B_0}{\sigma_\varepsilon} - \frac{x}{\sigma_\varepsilon}\right)}{\Psi\left(-\frac{\underline{v} - B_0}{\sigma_\varepsilon} - \frac{x}{\sigma_\varepsilon}\right) - \Psi\left(-\frac{\overline{v} - B_0}{\sigma_\varepsilon} - \frac{x}{\sigma_\varepsilon}\right)}$$

$$= \frac{\psi\left(\frac{\overline{v} - B_0}{\sigma_\varepsilon} - \frac{x}{\sigma_\varepsilon}\right) - \psi\left(\frac{\underline{v} - B_0}{\sigma_\varepsilon} - \frac{x}{\sigma_\varepsilon}\right)}{\Psi\left(\frac{\overline{v} - B_0}{\sigma_\varepsilon} - \frac{x}{\sigma_\varepsilon}\right) - \Psi\left(\frac{\underline{v} - B_0}{\sigma_\varepsilon} - \frac{x}{\sigma_\varepsilon}\right)}$$

$$= -T(x + B_0; \underline{v}, \overline{v}),$$

where the third and fourth equality follows from the condition $\overline{v} - B_0 = -(\underline{v} - B_0)$.

$$Premium_1(\underline{v}, \overline{v}) - Premium_0$$

$$= -E\left[\sigma_\varepsilon \cdot T(X; \underline{v}, \overline{v})\right]$$

$$= -\sigma_\varepsilon \int_{-\infty}^{+\infty} T(x; \underline{v}, \overline{v}) \frac{1}{\sqrt{2\pi}\sigma_X} e^{-\frac{(x - B_0)^2}{2\sigma_X^2}} dx$$

$$= -\sigma_\varepsilon \int_{-\infty}^{+\infty} T(x + B_0; \underline{v}, \overline{v}) \frac{1}{\sqrt{2\pi}\sigma_X} e^{-\frac{x^2}{2\sigma_X^2}} dx$$

$$= -\sigma_\varepsilon \left[ \int_0^{+\infty} T(x + B_0; \underline{v}, \overline{v}) \frac{1}{\sqrt{2\pi}\sigma_X} e^{-\frac{x^2}{2\sigma_X^2}} dx \right.$$

$$\left. + \int_{-\infty}^{0} T(x + B_0; \underline{v}, \overline{v}) \frac{1}{\sqrt{2\pi}\sigma_X} e^{-\frac{x^2}{2\sigma_X^2}} dx \right]$$

$$= -\sigma_\varepsilon \left[ \int_0^{+\infty} T(x + B_0; \underline{v}, \overline{v}) \frac{1}{\sqrt{2\pi}\sigma_X} e^{-\frac{x^2}{2\sigma_X^2}} dx \right.$$

$$\left. + \int_{0}^{+\infty} T(-x + B_0; \underline{v}, \overline{v}) \frac{1}{\sqrt{2\pi}\sigma_X} e^{-\frac{x^2}{2\sigma_X^2}} dx \right]$$

$$= -\sigma_\varepsilon \int_0^{+\infty} \left[ T(x + B_0; \underline{v}, \overline{v}) + T(-x + B_0; \underline{v}, \overline{v}) \right] \frac{1}{\sqrt{2\pi}\sigma_X} e^{-\frac{x^2}{2\sigma_X^2}} dx$$

$$= 0.$$

So, it is proved that $Premium_1(\underline{v}, \overline{v}) = Premium_0$ if $v_m = B_0$. If $v_m > B_0$, the interval $[\underline{v} - v_m + B_0,\ \overline{v} - v_m + B_0]$ whose midpoint is $B_0$ is lower than $[\underline{v},\ \overline{v}]$. Since the disclosure of a higher range causes a lower premium, we have

$$Premium_1(\underline{v},\ \overline{v}) < Premium_1(\underline{v} - v_m + B_0,\ \overline{v} - v_m + B_0) = Premium_0.$$

If $v_m < B_0$, the interval $[\underline{v} - v_m + B_0,\ \overline{v} - v_m + B_0]$ whose midpoint is $B_0$ is higher than $[\underline{v},\ \overline{v}]$. Then, we have

$$Premium_1(\underline{v},\ \overline{v}) > Premium_1(\underline{v} - v_m + B_0,\ \overline{v} - v_m + B_0) = Premium_0.$$

**A.13 The proof of Proposition 7**

By equation (25), $B_0 = \mu_0 + \Theta$, where

$$\Theta \triangleq -Z\gamma\sigma_\varepsilon^2 - \frac{Z\gamma^3 x_U \sigma_u^2 \sigma_\varepsilon^4 \sigma_y^2}{\gamma^2 \sigma_\varepsilon^4 \sigma_y^2 + x_I^2 \sigma_u^2 + \gamma^3 x_I \sigma_u^2 \sigma_\varepsilon^2 \sigma_y^2}.$$

Remind that $X \triangleq \tau u + \alpha y + \beta \sim N(\mu_0 + \Theta, \sigma_X^2)$ with $\sigma_X^2 = \tau^2\sigma_u^2 + \alpha^2\sigma_y^2$.

$$\lim_{D_{\overline{v}} \to +\infty} \frac{\partial Premium_1(\underline{v}, \overline{v})}{\partial \overline{v}}$$

$$= -\lim_{D_{\overline{v}} \to +\infty} E\left[ \frac{\partial \boldsymbol{J}_{[\underline{v},\ \overline{v}]}(\tau u + \alpha y + \beta)}{\partial \overline{v}} \right]$$

$$= -\lim_{D_{\overline{v}}\to+\infty} E\left[\frac{f_\varepsilon(\overline{v}-(\tau u+\alpha y+\beta))\int_{\underline{v}-(\tau u+\alpha y+\beta)}^{\overline{v}-(\tau u+\alpha y+\beta)}[\overline{v}-(\tau u+\alpha y+\beta)-x]f_\varepsilon(x)\,dx}{\left(\int_{\underline{v}-(\tau u+\alpha y+\beta)}^{\overline{v}-(\tau u+\alpha y+\beta)} f_\varepsilon(x)\,dx\right)^2}\right]$$

$$= -\lim_{D_{\overline{v}}\to+\infty} \int_{-\infty}^{+\infty} \frac{f_\varepsilon(\overline{v}-t)\int_{\underline{v}-t}^{\overline{v}-t}[\overline{v}-t-x]f_\varepsilon(x)\,dx}{\left(\int_{\underline{v}-t}^{\overline{v}-t} f_\varepsilon(x)\,dx\right)^2}\cdot\frac{1}{\sqrt{2\pi}\sigma_X}e^{-\frac{(t-\mu_0-\Theta)^2}{2\sigma_X^2}}\,dt$$

$$= -\lim_{D_{\overline{v}}\to+\infty} \int_{-\infty}^{+\infty} \frac{f_\varepsilon(\overline{v}-\mu_0-\Theta-t)\int_{\underline{v}-\mu_0-\Theta-t}^{\overline{v}-\mu_0-\Theta-t}[\overline{v}-\mu_0-\Theta-t-x]f_\varepsilon(x)\,dx}{\left(\int_{\underline{v}-\mu_0-\Theta-t}^{\overline{v}-\mu_0-\Theta-t} f_\varepsilon(x)\,dx\right)^2}$$

$$\cdot\frac{1}{\sqrt{2\pi}\sigma_X}e^{-\frac{t^2}{2\sigma_X^2}}\,dt$$

$$= -\lim_{D_{\overline{v}}\to+\infty} \int_{-\infty}^{+\infty} \frac{f_\varepsilon(D_{\overline{v}}-\Theta-t)\int_{\underline{v}-\mu_0-\Theta-t}^{D_{\overline{v}}-\Theta-t}[D_{\overline{v}}-\Theta-t-x]f_\varepsilon(x)\,dx}{\left(\int_{\underline{v}-\mu_0-\Theta-t}^{D_{\overline{v}}-\Theta-t} f_\varepsilon(x)\,dx\right)^2}$$

$$\cdot\frac{1}{\sqrt{2\pi}\sigma_X}e^{-\frac{t^2}{2\sigma_X^2}}\,dt$$

$$= -\int_{-\infty}^{+\infty} \lim_{D_{\overline{v}}\to+\infty} \frac{f_\varepsilon(D_{\overline{v}}-\Theta-t)\int_{\underline{v}-\mu_0-\Theta-t}^{D_{\overline{v}}-\Theta-t}[D_{\overline{v}}-\Theta-t-x]f_\varepsilon(x)\,dx}{\left(\int_{\underline{v}-\mu_0-\Theta-t}^{D_{\overline{v}}-\Theta-t} f_\varepsilon(x)\,dx\right)^2}$$

$$\cdot\frac{1}{\sqrt{2\pi}\sigma_X}e^{-\frac{t^2}{2\sigma_X^2}}\,dt.$$

For any $t \in R$,

$$0 \le \frac{f_\varepsilon(D_{\overline{v}}-\Theta-t)\int_{\underline{v}-\mu_0-\Theta-t}^{D_{\overline{v}}-\Theta-t}[D_{\overline{v}}-\Theta-t-x]f_\varepsilon(x)\,dx}{\left(\int_{\underline{v}-\mu_0-\Theta-t}^{D_{\overline{v}}-\Theta-t} f_\varepsilon(x)\,dx\right)^2}\cdot\frac{1}{\sqrt{2\pi}\sigma_X}e^{-\frac{t^2}{2\sigma_X^2}}$$

$$\le \frac{f_\varepsilon(D_{\overline{v}}-\Theta-t)\left(D_{\overline{v}}-\underline{v}+\mu_0\right)^2}{\left(\int_{\underline{v}-\mu_0-\Theta-t}^{D_{\overline{v}}-\Theta-t} f_\varepsilon(x)\,dx\right)^2}\cdot\frac{1}{2\pi\sigma_\varepsilon\sigma_X}e^{-\frac{t^2}{2\sigma_X^2}} \xrightarrow{\ when\ D_{\overline{v}}\to+\infty\ } 0.$$

So, we have

$$\lim_{D_{\overline{v}}\to+\infty} \frac{f_\varepsilon(D_{\overline{v}}-\Theta-t)\int_{\underline{v}-\mu_0-\Theta-t}^{D_{\overline{v}}-\Theta-t}[D_{\overline{v}}-\Theta-t-x]f_\varepsilon(x)\,dx}{\left(\int_{\underline{v}-\mu_0-\Theta-t}^{D_{\overline{v}}-\Theta-t} f_\varepsilon(x)\,dx\right)^2}\cdot\frac{1}{\sqrt{2\pi}\sigma_X}e^{-\frac{t^2}{2\sigma_X^2}} = 0$$

and then

$$\lim_{D_{\overline{v}} \to +\infty} \frac{\partial Premium_1(\underline{v}, \overline{v})}{\partial \overline{v}} = 0.$$

It can be proved that $\lim_{D_{\underline{v}} \to +\infty} \frac{\partial Premium_1(\underline{v}, \overline{v})}{\partial \underline{v}} = 0$ in a similar way.

### A.14 The proof of Proposition 8

$$\lim_{D \to +\infty} \frac{\partial Premium_1(\underline{v}, \overline{v})}{\partial \mu_0}$$

$$= -\lim_{D \to +\infty} \frac{\partial}{\partial \mu_0} E\left[\boldsymbol{J}_{[\underline{v},\ \overline{v}]}(\tau u + \alpha y + \beta)\right]$$

$$= -\lim_{D \to +\infty} \frac{\partial}{\partial \mu_0} \int_{-\infty}^{+\infty} \boldsymbol{J}_{[\underline{v},\ \overline{v}]}(x) \cdot \frac{1}{\sqrt{2\pi}\sigma_X} e^{-\frac{(x-\mu_0-\Theta)^2}{2\sigma_X^2}} dx$$

$$= -\lim_{D \to +\infty} \frac{\partial}{\partial \mu_0} \int_{-\infty}^{+\infty} \boldsymbol{J}_{[\underline{v},\ \overline{v}]}(x + \mu_0 + \Theta) \cdot \frac{1}{\sqrt{2\pi}\sigma_X} e^{-\frac{x^2}{2\sigma_X^2}} dx$$

$$= -\lim_{D \to +\infty} \int_{-\infty}^{+\infty} \frac{\partial}{\partial \mu_0} \left[\boldsymbol{J}_{[\underline{v},\ \overline{v}]}(x + \mu_0 + \Theta) \cdot \frac{1}{\sqrt{2\pi}\sigma_X} e^{-\frac{x^2}{2\sigma_X^2}}\right] dx$$

$$= -\lim_{D \to +\infty} \int_{-\infty}^{+\infty} \boldsymbol{H}_{[\underline{v},\ \overline{v}]}(x + \mu_0 + \Theta) \cdot \frac{1}{\sqrt{2\pi}\sigma_X} e^{-\frac{x^2}{2\sigma_X^2}} dx$$

$$= -\int_{-\infty}^{+\infty} \lim_{D \to +\infty} \boldsymbol{H}_{[\underline{v},\ \overline{v}]}(x + \mu_0 + \Theta) \cdot \frac{1}{\sqrt{2\pi}\sigma_X} e^{-\frac{x^2}{2\sigma_X^2}} dx$$

For any $x \in R$, if $\mu_0 > \overline{v}$, we have

$$\begin{aligned}
&\lim_{D \to +\infty} \boldsymbol{H}_{[\underline{v},\ \overline{v}]}(x + \mu_0 + \Theta) \\
&= \lim_{D \to +\infty} \boldsymbol{H}_{[-L,0]}(x + \mu_0 - \overline{v} + \Theta) \\
&= \lim_{D \to +\infty} \boldsymbol{H}_{[-L,0]}(x + D + \Theta) \\
&= 0,
\end{aligned}$$

where the first and third equality follow from the item (5) and (4) in appendix A.6. Similarly, if $\mu_0 < \underline{v}$, then

$$\begin{aligned}
&\lim_{D \to +\infty} \boldsymbol{H}_{[\underline{v},\ \overline{v}]}(x + \mu_0 + \Theta) \\
&= \lim_{D \to +\infty} \boldsymbol{H}_{[0,\mathrm{L}]}\left(x + \mu_0 - \underline{v} + \Theta\right) \\
&= \lim_{D \to +\infty} \boldsymbol{H}_{[0,\mathrm{L}]}(x - D + \Theta) \\
&= 0.
\end{aligned}$$

As a result, we derive $\lim_{D \to +\infty} \boldsymbol{H}_{[\underline{v},\ \overline{v}]}(x + \mu_0 + \Theta) = 0$ and it follows that

$$\lim_{D \to +\infty} \frac{\partial Premium_1(\underline{v}, \overline{v})}{\partial \mu_0} = 0.$$

For the sensitivity of asset premium to movement of the range $[\underline{v},\ \overline{v}]$, we have

$$\begin{aligned}
&\lim_{D \to +\infty} \frac{\partial Premium_1(\underline{v}, \overline{v})}{\partial v_m} \\
&= -\lim_{D \to +\infty} \frac{\partial}{\partial v_m} E\left[\boldsymbol{J}_{[\underline{v},\ \overline{v}]}(\tau u + \alpha y + \beta)\right] \\
&= -\lim_{D \to +\infty} \frac{\partial}{\partial v_m} \int_{-\infty}^{+\infty} \boldsymbol{J}_{[\underline{v},\ \overline{v}]}(x) \cdot \frac{1}{\sqrt{2\pi}\sigma_X} e^{-\frac{(x-\mu_0-\Theta)^2}{2\sigma_X^2}} dx \\
&= -\lim_{D \to +\infty} \frac{\partial}{\partial v_m} \int_{-\infty}^{+\infty} \boldsymbol{J}_{[\underline{v},\ \overline{v}]}(x + \mu_0 + \Theta) \cdot \frac{1}{\sqrt{2\pi}\sigma_X} e^{-\frac{x^2}{2\sigma_X^2}} dx \\
&= -\lim_{D \to +\infty} \int_{-\infty}^{+\infty} \frac{\partial}{\partial v_m} \left[\boldsymbol{J}_{[\underline{v},\ \overline{v}]}(x + \mu_0 + \Theta) \cdot \frac{1}{\sqrt{2\pi}\sigma_X} e^{-\frac{x^2}{2\sigma_X^2}}\right] dx \\
&= -\lim_{D \to +\infty} \int_{-\infty}^{+\infty} \left(1 - \boldsymbol{H}_{[\underline{v},\ \overline{v}]}(x + \mu_0 + \Theta)\right) \cdot \frac{1}{\sqrt{2\pi}\sigma_X} e^{-\frac{x^2}{2\sigma_X^2}} dx \\
&= -\int_{-\infty}^{+\infty} \lim_{D \to +\infty} \left(1 - \boldsymbol{H}_{[\underline{v},\ \overline{v}]}(x + \mu_0 + \Theta)\right) \cdot \frac{1}{\sqrt{2\pi}\sigma_X} e^{-\frac{x^2}{2\sigma_X^2}} dx \\
&= -\int_{-\infty}^{+\infty} \frac{1}{\sqrt{2\pi}\sigma_X} e^{-\frac{x^2}{2\sigma_X^2}} dx \\
&= -1,
\end{aligned}$$

where the fifth equality follows from Lemma 3.

### A.15 The proof of Proposition 9

$$\begin{aligned}
\lim_{\substack{\overline{v} \to +\infty \\ \underline{v} \to -\infty}} Premium_1(\underline{v}, \overline{v}) &= \mu_0 - \lim_{\substack{\overline{v} \to +\infty \\ \underline{v} \to -\infty}} E\left[\boldsymbol{J}_{[\underline{v},\ \overline{v}]}(\tau u + \alpha y + \beta)\right] \\
&= \mu_0 - E\left[\lim_{\substack{\overline{v} \to +\infty \\ \underline{v} \to -\infty}} \boldsymbol{J}_{[\underline{v},\ \overline{v}]}(\tau u + \alpha y + \beta)\right] \\
&= \mu_0 - E[p_0] \\
&= Premium_0(\underline{v}, \overline{v}).
\end{aligned}$$

Additionally,

$$
\begin{aligned}
\lim_{\substack{\overline{v}\to+\infty\\ \underline{v}\to-\infty}} \frac{\partial Premium_1(\underline{v},\overline{v})}{\partial v_m} &= -\lim_{\substack{\overline{v}\to+\infty\\ \underline{v}\to-\infty}} \frac{\partial}{\partial v_m} E\left[\boldsymbol{J}_{[\underline{v},\ \overline{v}]}(\tau u+\alpha y+\beta)\right] \\
&= -\lim_{\substack{\overline{v}\to+\infty\\ \underline{v}\to-\infty}} E\left[\frac{\partial}{\partial v_m}\boldsymbol{J}_{[\underline{v},\ \overline{v}]}(\tau u+\alpha y+\beta)\right] \\
&= -\lim_{\substack{\overline{v}\to+\infty\\ \underline{v}\to-\infty}} E\left[1-\boldsymbol{H}_{[\underline{v},\ \overline{v}]}(\tau\tilde{u}+\alpha\tilde{y}+\beta)\right] \\
&= -1+E\left[\lim_{\substack{\overline{v}\to+\infty\\ \underline{v}\to-\infty}} \boldsymbol{H}_{[\underline{v},\ \overline{v}]}(\tau\tilde{u}+\alpha\tilde{y}+\beta)\right] \\
&= 0.
\end{aligned}
$$

The third and fifth equality follow from Lemma 3 and the item (3) in appendix A.6 respectively.

**Author Contributions:** Conceptualization, J.S.; methodology, J.S.; software, J.S.; validation, J.S.; formal analysis, J.S.; investigation,J.S.; resources, J.S.; data curation, J.S.; writing—original draft preparation, J.S.; writing—review and editing, Y.Z.; visualization, J.S.; supervision, Y.Z.; project administration, J.S.; funding acquisition, J.S.. All authors have read and agreed to the published version of the manuscript.

**Data Availability Statement:** The raw data supporting the conclusions of this article will be made available by the authors on request.

**Conflicts of Interest:** There is no conflicts of interest.

**References**

1. Grossman, S. J.; Stiglitz, J. E. On the impossibility of informationally efficient markets. The American economic review 1980, 70, 393-408.
2. Epstein, L. G.; Schneider, M. Ambiguity, information quality, and asset pricing. The Journal of Finance 2008, 63, 197-228.
3. Mele, A.; Sangiorgi, F. Uncertainty, information acquisition, and price swings in asset markets. The Review of Economic Studies 2015, 82, 1533-1567.
4. Hahn, G.; Kwon, J. Y. How Do Ambiguity and Risk Aversion Affect Price Volatility under Asymmetric Information? Asia-Pacific Journal of Financial Studies 2015, 44, 616-634.

5. Condie, S.; Ganguli, J. The pricing effects of ambiguous private information. Journal of Economic Theory 2017, 172, 512-557.
6. Huo, Z.; Pedroni, M.; Pei, G. Bias and sensitivity under ambiguity. American Economic Review 2024, 114, 4091-4133.
7. Easley, D.; O'Hara, M.; Yang, L. Opaque trading, disclosure, and asset prices: Implications for hedge fund regulation. The Review of Financial Studies 2014, 27, 1190-1237.
8. Huang, H. H.; Zhang, S.; Zhu, W. Limited participation under ambiguity of correlation. Journal of Financial Markets 2017, 32, 97-143.
9. Illeditsch, P. K.; Ganguli, J. V.; Condie, S. Information inertia. The Journal of Finance 2021, 76, 443-479.
10. Anthropelos, M.; Schneider, P. Optimal investment and equilibrium pricing under ambiguity. Review of Finance 2024, 28, 1759-1805.
11. Ghazi, S.; Schneider, M.; Strauss, J. Market Ambiguity Attitude Restores the Risk-Return Trade-Off. Management Science, 2025, online.
12. Kondor, P. The more we know about the fundamental, the less we agree on the price. The Review of Economic Studies 2012, 79, 1175-1207.
13. Goldstein, I.; Yang, L. Information diversity and complementarities in trading and information acquisition. The Journal of Finance 2015, 70, 1723-1765.
14. Liu, X.; Huang, W.; Liu, B.; Zhang, X. Strategic leakage of private information. The North American Journal of Economics and Finance 2019, 47, 637-644.
15. Yang, L.; Zhu, H. Back-running: Seeking and hiding fundamental information in order flows. The Review of Financial Studies 2020, 33, 1484-1533.
16. Dumas, B.; Lewis, K. K.; Osambela, E. Differences of opinion and international equity markets. The Review of Financial Studies 2017, 30, 750-800.
17. Banerjee, S.; Davis, J.; Gondhi, N. When transparency improves, must prices reflect fundamentals better? The Review of Financial Studies 2018, 31, 2377-2414.
18. Liu, X.; Liu, S.; Lu, L.; Xiong, X. Voluntary information disclosure with heterogeneous beliefs. Journal of Economic Dynamics and Control, 2021, 124, 104081.
19. Hu, D.; Wang, H. Heterogeneous beliefs with preference interdependence and asset pricing. International Review of Economics & Finance 2024, 93, 1-37.
20. Mondria, J.; Vives, X.; Yang, L. Costly interpretation of asset prices. Management Science 2022, 68, 52-74.
21. Huang, S.; Qiu, Z.; Yang, L. Institutionalization, delegation, and asset prices. Journal of Economic Theory 2020, 186, 104977.

22. Goldstein, I.; Yang, L. Information disclosure in financial markets. Annual Review of Financial Economics 2017, 9, 101-125.

23. Easley, D.; O'Hara, M. Ambiguity and nonparticipation: The role of regulation. The Review of Financial Studies 2009, 22, 1817-1843.

24. Kodres, L. E.; Pritsker, M. A. rational expectations model of financial contagion. The journal of finance 2002, 57, 769-799.